\documentclass[aps,prl,twocolumn,superscriptaddress]{revtex4-2}

\usepackage[T1]{fontenc}
\usepackage{graphicx,subfigure}
\usepackage{amsmath,amssymb,bm,amsthm}
\usepackage{mathtools}
\usepackage{multirow}
\usepackage{textcomp}
\usepackage{float}
\usepackage[dvipsnames]{xcolor}
\usepackage[normalem]{ulem}
\usepackage{dsfont}
\usepackage{url}
\usepackage{mathrsfs}
\usepackage{lipsum}
\usepackage{tikz}
\usepackage{enumerate}

\usetikzlibrary{shapes, arrows, positioning}
\usepackage[colorlinks=true, allcolors=blue]{hyperref}

\newcommand{\avg}[1]{\left\langle #1 \right\rangle}
\newcommand{\ket}[1]{\left| #1 \right\rangle}
\newcommand{\bra}[1]{\left\langle #1 \right|}

\newcommand{\epsd}{\varepsilon_{\text{d}}}
\newcommand{\Bb}{\mathbf{B}}
\newcommand{\bb}{\mathbf{b}}
\newcommand{\Gb}{\mathbf{G}}
\newcommand{\Qb}{\mathbf{Q}}
\newcommand{\betab}{\boldsymbol{\beta}}
\newcommand{\gammab}{\boldsymbol{\gamma}}
\newtheorem{theorem}{Theorem}

\begin{document}

\title{Generalized Quantum Repeater Graph States}

\author{Bikun Li}
\email{bikunli@uchicago.edu}
\affiliation{Pritzker School of Molecular Engineering, University of Chicago, Chicago, Illinois 60637, USA}
\author{Kenneth Goodenough}
\email{kgoodenough@cics.umass.edu}
\affiliation{College of Information and Computer Sciences, University of Massachusetts Amherst, Amherst, Massachusetts 01003, USA}
\author{Filip Rozp\k{e}dek}
\email{frozpedek@cics.umass.edu}
\affiliation{College of Information and Computer Sciences, University of Massachusetts Amherst, Amherst, Massachusetts 01003, USA}
\author{Liang Jiang}
\email{liangjiang@uchicago.edu}
\affiliation{Pritzker School of Molecular Engineering, University of Chicago, Chicago, Illinois 60637, USA}

\date{\today}

\begin{abstract}
    All-photonic quantum repeaters are essential for establishing long-range quantum entanglement. 
    Within repeater nodes, reliably performing entanglement swapping is a key component of scalable quantum communication. 
    To tackle the challenge of probabilistic Bell state measurement in linear optics, which often leads to information loss, various approaches have been proposed to ensure the loss tolerance of distributing a single ebit. 
    We have generalized previous work regarding repeater graph states with elaborate connectivity, enabling the efficient establishment of exploitable ebits at a finite rate with high probability.
    We demonstrate that our new scheme significantly outperforms the previous work with much flexibility and discuss the generation overhead of such resource states.
    These findings offer new insights into the scalability and reliability of loss-tolerant quantum networks.
\end{abstract}

\maketitle

\textit{Introduction} -- 
Quantum entanglement is one of the key resources that differentiates quantum information technology from classical information technology. 
The fundamental unit of bipartite entanglement is known as an ebit. 
Reliably distributing abundant ebits across various parties allows us to establish a network of quantum entanglement. 
Such a quantum network is imperative for transferring quantum information between distant nodes, facilitating many novel applications in quantum communication, distributed quantum computation, and quantum sensing ~\cite{bartolucci2021fusionbased, Kitaev2003, Briegel, Sangouard2011, Azuma2015, Damian2020, E92}.
Since classical information can be efficiently delivered by optical signals, the most common way of preparing long-range entangled quantum states is by using photonic qubits. 
To address the ubiquitous noise in transmission lines, quantum repeater schemes~\cite{Muralidharan2016, RevModPhys.95.045006, Munro2015review, PhysRevResearch.5.043056} have been proposed to mitigate the photon loss and Pauli errors in the quantum channel.
Similar to its classical counterpart, the long distance of quantum information transmission is divided into shorter segments. 
The error rate in these segments is therefore sufficiently small to be managed by unique quantum techniques. 
Bridging all these segments in series with repeater nodes yields a better entanglement generation rate than direct transmission.
Another advantage of the repeater scheme is that short-range entangled states can be prepared in parallel, allowing distant ends to process quantum information without long waiting times.

All-photonic repeater schemes~\cite{Azuma2015,PhysRevA.95.012304,Ewert2014PRL,Ewert2016PRL, fukui2021all,niu2022all,rozpkedek2023all,patil2024improved} are usually realized by preparing quantum resource states, which are sent to the nearest repeater stations, and processed by linear optics approaches. 
For discrete-variable all-photonic schemes, the crucial part, entanglement swapping, is implemented by nondeterministic fusion operations within each repeater station, requiring only polarizing beam splitters and single-photon detectors \cite{fusiongates}. 
The photonic resource state is engineered in a special way to overcome both the channel losses and the probabilistic nature of optical Bell state measurements \cite{nogo_deterministic_BSM}. 
Measurement-based repeater schemes based on quantum resource states have been proposed in early works \cite{Zwerger2012PRA, Zwerger2018PRL}. However, their model becomes inefficient if the Bell state measurement (BSM) has a high rate of loss. Particularly, the Choi-state-based approach in~\cite{Zwerger2018PRL} fails if the loss of BSM exceeds $50\%$, since the no-cloning theorem implies that the quantum channel represented by the Choi state has a vanishing quantum capacity~\cite{ErasChnCapacity1997PRL}. 

Although there have been some promising techniques of increasing the BSM success rate \cite{Ewert2014PRL, Ewert2016PRL}, a prominent instance of a resource state for quantum repeaters is the \emph{repeater graph state} (RGS)~\cite{Azuma2015, Li2019}. The RGS adapts to any BSM loss by rerouting the successful fusion operations with a special quantum graph state.
The RGS can counter the logical failure of fusion operations by scaling up its size. Recent works~\cite{ButerakosPRX2017, Zhan2023performanceanalysis, PhysRevA.95.012304,patil2024improved} have discussed the efficient implementation and robustness of this scheme. 
Nevertheless, at most a single ebit will be distributed once any of these protocols succeeds. 
In other words, they have a vanishing \emph{rate} --- the number of ebits that can be established per photon used in the RGS goes to zero. These prior works based on \cite{Azuma2015} all suffered from low rate per run of the protocol, due to the inefficient BSM rerouting.

In this Letter, we address the above issue by modifying and generalizing the previous RGS with improved graphs. These graph states are optimized and flexible, in the sense that they can be derived from any good classical binary error-correcting codes. 
In other words, assuming perfect operations except for lossy BSMs, the ratio between the number of extracted ebits and the number of qubits used can be maximized by using repeater graph states corresponding to capacity-achieving error-correcting codes~\cite{MacKay2003, Luby2001EECC, RaptorCodes}. 
We also analyze the minimum overhead of preparing this new class of RGS in the deterministic generation scheme. The significant outperformance of the ebit rate by our scheme is demonstrated by an example of a quantum key distribution (QKD) protocol.

\begin{figure}[t]
    \centering
    \includegraphics[width=1.0\linewidth]{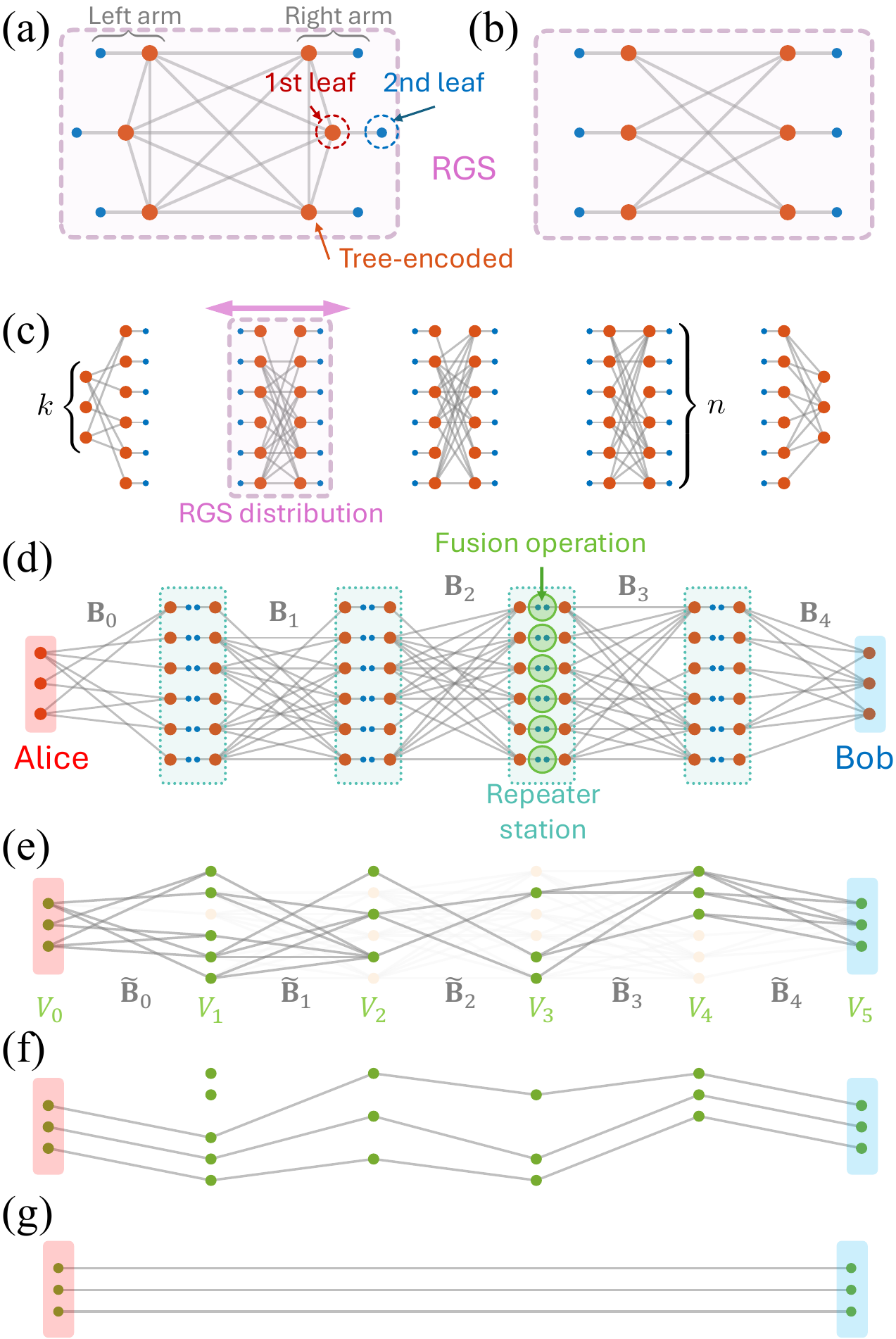}
    \caption{(a) An RGS example $|\overline{\mathcal{G}}_c^3\rangle$ given by \cite{Azuma2015} (purple box). 
    (b) A modified version of (a) with a bipartite complete graph. 
    (c) An instance of distributing a generalized RGS (purple box) to neighboring repeater stations (green boxes in (d)). (d) Each RGS is defined by a biadjacency matrix $\Bb_{a}$. The distant parties are labeled as Alice (red) and Bob (blue). Incoming qubits are collected and processed in the repeater station. In this case, $N_R = 4$, $k=3$, and $n=6$. 
    (e) A possible intermediate LTG state given by approximately $50\%$ successful fusion operations. The faded background shows the otherwise trellis graph if all fusions are successful. 
    (f) A local-CNOT equivalent graph state with respect to the graph state in (e), which implies $k$ linear cluster states are encoded at the logical level. 
    Panel (g) shows the final state $\ket{\phi_{AB}}$ obtained from (e). 
    }
    \label{fig:rgs}
\end{figure}

\textit{RGS scheme and improvement} --
Let's briefly review the concept of the quantum repeater graph state (RGS) as introduced in \cite{Azuma2015}. 
A graph state $\ket{\mathcal{G}}$ of qubits is a pure stabilizer state defined by a simple graph $\mathcal{G} = (V, E)$, where $V$ is the set of vertices or qubits and $E$ is the set of edges. 
The state $\ket{\mathcal{G}}$ is defined as the common eigenvector of the stabilizer generators $\hat{K}_i := X_i\prod_{j\in\mathcal{N}(i)}Z_j, \forall i \in V$. Here $\mathcal{N}(i):=\{j:\{i,j\}\in E\}$ is the set of vertex $i$'s neighbors, and $X$ and $Z$ are Pauli operators.
The scheme in \cite{Azuma2015} utilizes a graph $\mathcal{G}_c^n$ consisting of $4n$ vertices, which includes $2n$ inner `1st leaves' and $2n$ outer `2nd leaves' (see Fig.~\ref{fig:rgs}(a)). 
The inner leaves are fully connected as a complete graph. The edges between the 1st and 2nd leaves can be viewed as the ``arms'' of an RGS. 
Both leaves are sent through the transmission channel, in which the 1st leaf qubits are encoded with a loss-resilient code to boost fidelity in the short-range quantum state distributions in each transmission segment.
A common choice for such a loss-resilient code is the so-called tree code \cite{Varnava2006}. 
The entire graph with tree-encoding is denoted as $\overline{\mathcal{G}}_c^n$. 
To perform the repeater protocol, the repeater station first attempts to perform $n$ BSMs via fusion operations on the adjacent 2nd leaves of two incoming RGSs~\cite{fusiongates, SuppMat}. 
Denote the total failure rate of the BSM and the photon loss as $p_{f}$ (typically, $p_f\ge 0.5$).
The chance of having at least one successful BSM is $1-p_{f}^n$ since at most $n-1$ failed fusion operations can be tolerated in each repeater station.
Once a BSM succeeds, the adjacent qubit on the 1st leaf is measured in the logical $X$ basis, and the other qubits of the 1st leaf are pruned using logical $Z$ measurements. 
In tree encodings, the logical $X$ and $Z$ measurements are performed by single-qubit measurements and data postprocessing. 
Note that the intermediate state is always equivalent to some graph state. 
The graphical transformation \cite{Hein2004PRA} and the full connectivity of each complete graph ensure that, with high probability, a linear cluster state (the path graph) connecting Alice and Bob shows up as an intermediate state, which can then be used to establish at most one ebit between Alice and Bob~\cite{Briegel}.

We will present what is \textit{necessary} to generalize the original RGS. Note that the qubit number overhead of the above RGS per final ebit is proportional to $n$, which leads to a vanishing encoding rate of ebits. 
This drawback stems from the fact that the bipartite entanglement entropy of the complete graph state is always $1$. 
The number of underlying ebits across the bipartition is given by the rank of the \textit{biadjacency matrix} \cite{Hein2004PRA}.
This is because local Clifford unitaries (with respect to the bipartition) can transform the state into decoupled ebits by diagonalizing the biadjacency matrix and disconnecting the edges within each party. 
We should highlight that using multiple copies of $|\overline{\mathcal{G}}^n_c\rangle$ in parallel is convenient but still suboptimal, as the redundancy of each copy could have been shared to further suppress losses. 
Therefore, the optimized graph must possess elaborate coupling among ebits.

\begin{figure}[t]
    \centering
    \includegraphics[width=1.0\linewidth]{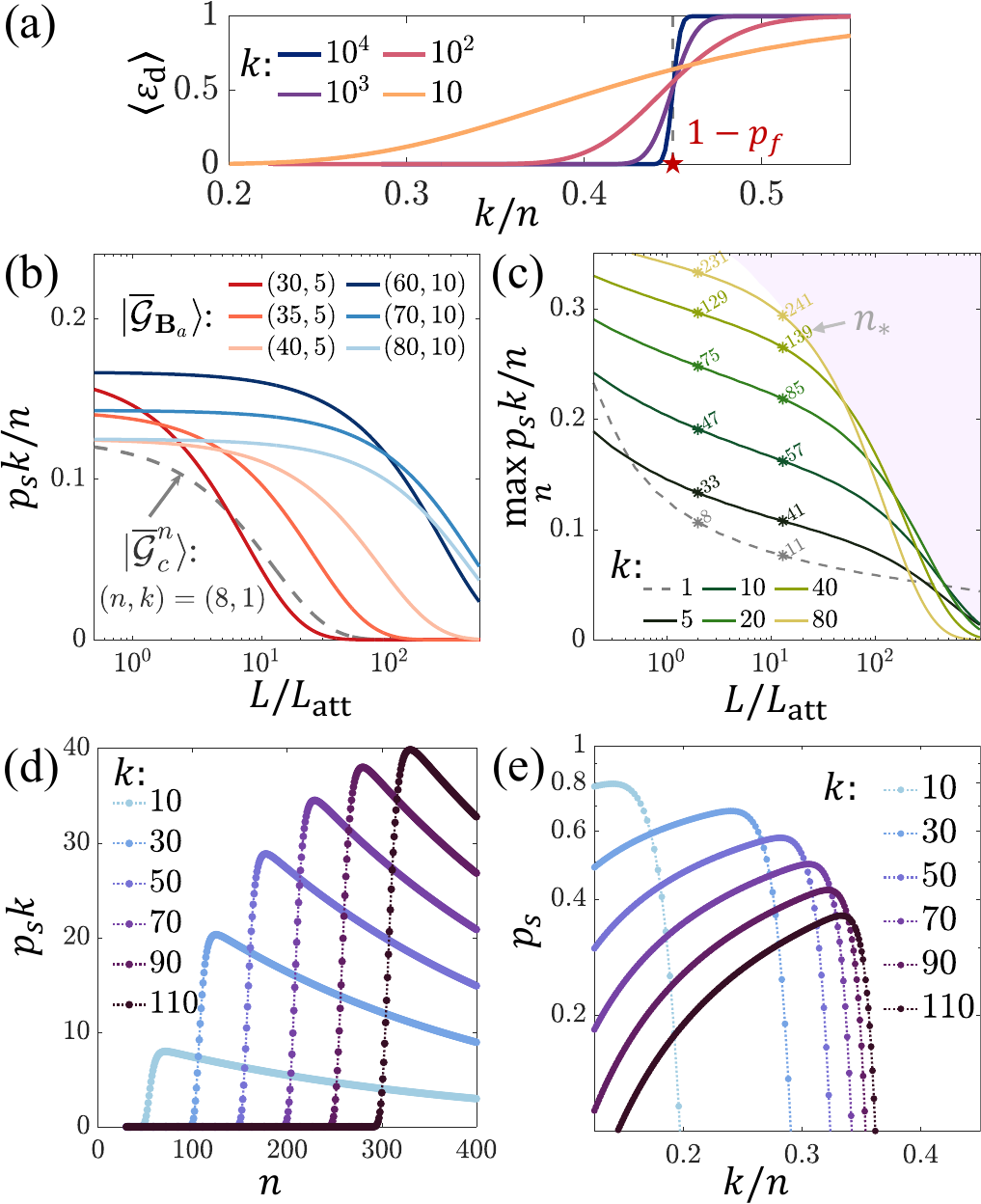}
    \caption{(a) The rank-deficient probability $\avg{\epsd}$ of the random submatrix $\widetilde{\Gb}$, where we have assumed $p_f = 0.55$.
    (b) The expected performance $k/n$ per run of the QKD protocol across the total distance $L$.  Further numerical optimization over the value of $n$ is presented in (c), where $n_*$ is the solution at the corresponding $L$. 
    The gray dashed curves in panels (b) and (c) represent the protocol in \cite{Azuma2015}, and our new protocols with generalized RGSs (random codes) are represented by solid curves.
    The branch vector of the tree encoding is set as $\bb = (5,11,4)$ for panels (b) to (e). 
    By fixing $L = 100L_{\mathrm{att}}$, panel (d) displays the expected number of ebits being estabilished with respect to different choices of $(n,k)$. Panel (e) shows the value of protocol success probability $p_s$ with respect to different $k/n$ values.
    }
    \label{fig:curves}
\end{figure}

The first step toward a better design is removing some edges in the RGS.
Previous works \cite{Tzitrin2018PRA, Russo2018PRB} have argued that the modified RGS in Fig.~\ref{fig:rgs}(b) does not hinder the rerouting of successful fusion operations. The key insight is that the biadjacency matrix $\Bb\in\mathbb{F}_2^{n\times n}$ of the bipartite complete graph can be decomposed as $\Bb = \Gb^T_L \Gb_R$, with $\Gb_{L/R} = (1,1,\cdots,1)$. 
In the original RGS protocol, the successful BSMs from the left (L) and right (R) arms of the RGS induce the trivial submatrix $\widetilde{\Gb}_{L/R}=1$ from the corresponding columns of $\Gb_{L/R}$. The resulting graph state is dictated by a biadjacency submatrix $\widetilde{\Bb}=\widetilde{\Gb}_{L}^T \widetilde{\Gb}_{R}=1$ with unit rank.
If we expect to establish more ebits at once, $\widetilde{\Bb}$ need to have some rank $k>1$. It happens as more columns of $\Gb_{L/R}$ are kept (corresponding to more fusions succeeding), the resulting $\widetilde{\Gb}_{L/R}$ is more likely to have rank $k>1$ (here and throughout this Letter, $\widetilde{\mathbf{A}}$ represents a submatrix of $\mathbf{A}$, defined by a random subset of $\mathbf{A}$'s columns).
This is possible if $\Gb$ has at least $k$ independent rows.
Heuristically, this feature is found in the generator matrix of erasure codes for classical bits. If $\Gb\in\mathbb{F}_2^{k\times n}$ is a generator matrix, which encodes $k$ information bits as $n$ variable bits, then $\widetilde{\Gb}_{L/R}$ is associated with the received variable bits after the bit erasure, which must have a full row rank to recover all $k$ information bits. From this perspective, the bipartite complete graph in Fig.~\ref{fig:rgs} is given by the repetition code with a poor rate, which implies that we can consider general erasure codes for better rates.
The maximal rate of perfectly transmitting information through the binary erasure channel is given by the capacity $\mathcal{C}_{e}(p_e) = 1 - p_e$ \cite{ErasChnCapacity1997PRL}, where $p_e$ ($0\le p_e \le 1$) is the erasure rate, which is equivalent to $p_f$ in our case. 
A capacity-approaching erasure code can protect the information from bit loss with a very small or even vanishing fraction of redundancy. 
In other words, given such a good code, $k$ bits of information can be recovered from \textit{arbitrary} $\widetilde{n}=k+\delta(k,\epsd)$ received bits with any recovery failure rate $\epsd>0$, where the redundancy fulfills $\lim_{k\to\infty}\delta(k,\epsd)/k = 0$.
Note that a random generator matrix uniformly sampled from $\mathbb{F}_2^{k\times n}$ already achieves the goal with $\mathcal{O}(k^3)$ decoding complexity \cite{Luby2001EECC,SuppMat}. The probability of obtaining a rank-deficient $\widetilde{\Gb}_{L/R}$ from $\Gb$ is the mean value $\avg{\epsd}$, which can be made exponentially small under some threshold (Fig.~\ref{fig:curves}(a)). We emphasize that the above inefficient decoding complexity can be improved to $\mathcal{O}(k)$ if we adopt modern practical erasure codes such as fountain codes and low-density parity-check codes~\cite{Gallager_LDPC, Luby2001EECC, RaptorCodes}.

\textit{Generalized RGS} --
The above discussion only ensures that many ebits may be established but has yet to explain how to sort out each Bell pair for practical use. 
We will demonstrate an erasure-code-inspired design that is \textit{sufficient} for establishing exploitable Bell pairs for Alice and Bob. 
Remarkably, only single-qubit measurements and postprocessing of measurement data are needed to extract $k$ Bell states in the end.
Our scheme is specified as follows, assuming there are $N_R>0$ repeater nodes. 
As shown in Fig.~\ref{fig:rgs}(c,d), an RGS with a biadjacency matrix $\Bb_{a}:=\Gb_{a}^{T}\Gb_{a+1}$ is prepared and distributed to neighboring repeater nodes $a$ and $a+1$, where $\Gb_{a}\in\mathbb{F}_2^{k\times n}$ is a generator matrix of an erasure code. $n(1-p_f) > k$ is required so the rate $k/n$ does not exceed the capacity $\mathcal{C}_e(p_f)$.
Similar to the complete graph RGS, the graph for the (encoded-) RGS is denoted as ($\overline{\mathcal{G}}_{\Bb_{a}}$) $\mathcal{G}_{\Bb_{a}}$. 
To further simplify the notation, the distant parties Alice and Bob can be treated as “boundary repeater nodes” $a = 0$ and $a = N_R + 1$ with boundary condition $\Gb_{0}=\Gb_{N_R+1} = \mathbf{I}_k$, where $\mathbf{I}_k$ is a $k\times k$ identity matrix. 
That is, $|\overline{\mathcal{G}}_{\Bb_{0}}\rangle$ and $|\overline{\mathcal{G}}_{\Bb_{N_R}}\rangle$ are distributed to Alice, Bob, and their adjacent repeater stations.

As in \cite{Azuma2015}, one can perform the BSM on qubits of the 2nd leaf by the fusion operation \cite{fusiongates, SuppMat}. 
In contrast to the original protocol, where only one successful BSM is kept, we will keep as many successful BSMs as possible and denote this number as $\widetilde{n}$. One of the two qubits of the corresponding 1st leaves is further measured in the $X$ basis whenever a successful BSM occurs \cite{SuppMat}.
The resulting graph state can be represented by a \textit{linear trellis graph} (LTG) as shown in Fig.~\ref{fig:rgs}(e), which is defined by a sequence of submatrices: $\widetilde{\Bb}_{a}=\widetilde{\Gb}_{a}^T\widetilde{\Gb}_{a+1}$.
The repeater station $a$ ($1\le a \le N_R$) further measures the green nodes in the $X$ basis, with the measurement outcome $\tilde{\mathbf{x}}_{a}\in \mathbb{F}_2^{\widetilde{n}}$. 
A remarkable fact is that the above single-qubit measurements effectively implement logical $X$ measurements for the quantum teleportation in Fig.~\ref{fig:rgs}(f) (also see Theorem~\ref{thm:local_CNOT_equiv}). The bit string $\mathbf{m}_{a}\in\mathbb{F}_2^k$ of these logical $X$ measurements can be solved from the equation $\widetilde{\mathbf{x}}_{a}=\mathbf{m}_{a}\widetilde{\Gb}_{a}$, which can be done in $\mathcal{O}(k)$ steps if a good code is used. 
We highlight that the measurement of $\widetilde{\mathbf{x}}_{a}$ can be further made fault tolerant by a good choice of $\Gb_a$ to counter both bit loss and bit flip \cite{SuppMat}.
Finally, Alice and Bob can extract $k$ Bell states from $k$ linear cluster states by the decoded $\mathbf{m}_{a}$. 
Let $\ket{\Phi^+}_{AB}:=\frac{1}{\sqrt{2}}(\ket{00}_{AB}+\ket{11}_{AB})$  be the Bell state between Alice and Bob, and the state shared by Alice and Bob after the repeaters' measurement is denoted as $\ket{\phi_{AB}}$ (Fig.~\ref{fig:rgs}(g)).
If $N_R$ is odd, then
\begin{equation}
    \ket{\phi_{AB}} = \bigotimes_{i=1}^k \prod_{\substack{\text{ even }a\\{\text{odd } b}}}Z_{A,i}^{(\mathbf{m}_{a})_i}X_{A,i}^{(\mathbf{m}_{b})_i}\ket{\Phi^+}_{AB,i}\;,
\end{equation}
where $(\mathbf{m}_{a})_i$ is the $i$th entry of $\mathbf{m}_{a}$, and the operator $U_{A,i}$ acts on the $i$th qubit of Alice. 
The dummy indices $a$ and $b$ are taken from $\left\{1,2,\cdots,N_R\right\}$. 
If $N_R$ is even, then additional Hadamard gates are applied to the above state $\ket{\phi_{AB}} \to \bigotimes_{i=1}^k H_{A,i}\ket{\phi_{AB}}$.

\textit{Transversal measurements} --
We now exhibit more rigorous details about the trellis graph and the transversal measurement of $\mathbf{m}_{a}$.
We can formally define the LTG $\mathcal{G}_{\text{tr}}:=(\bigcup_{a}V_{a},E_{\text{tr}})$ by disjoint subsets of vertices $V_{a}$ (for example, Fig.~\ref{fig:rgs}(e)). 
Any edge in an LTG can only connect adjacent $V_{a}$ and $V_{a+1}$ by a biadjacency matrix $\betab_{a}\in\mathbb{F}_2^{|V_a|\times |V_{a+1}|}$.
That is, under some graph isomorphism, an LTG can be defined by a sequence of biadjacency matrices $\betab_{a}$.
We also define the \textit{local-CNOT} operation associated with $\{V_a\}$ by arbitrary controlled-NOT gates locally acting on qubits within each $V_{a}$. 
The local-CNOT equivalence of two states means such operations can convert one state to another (Fig.~\ref{fig:rgs}(e,f)), which fulfills the following theorem:
\begin{theorem}\label{thm:local_CNOT_equiv}

    Let $\mathcal{G}_{\mathrm{tr}}$ be an LTG associated with $\{V_a\}$ and $\betab_{a} = \gammab_{a}^T\gammab_{a+1}$, where all $\gammab_{a}\in\mathbb{F}_2^{k\times |V_{a}|}$ have full row rank. Then the graph state $\ket{\mathcal{G}_{\mathrm{tr}}}$ is local-CNOT equivalent to another graph state $\ket{\mathcal{G}'_{\mathrm{tr}}}$ of an LTG $\mathcal{G}'_{\mathrm{tr}}$ with $\{V_a\}$ and $\betab'_{a} = \gammab'^{T}_{a} \gammab'_{a+1}$, where $\gammab'_{a}=[\mathbf{I}_k,\mathbf{0}_{k,|V_a|-k}]$ and $\mathbf{0}_{k,m}$ is a $k\times m$ zero block. 
\end{theorem}
The proof and examples are provided in \cite{SuppMat}.
Since the graph $\mathcal{G}'_{\text{tr}}$ consists of $k$ disjoint path graphs and some disconnected vertices, 
Theorem~\ref{thm:local_CNOT_equiv} tells us the ``local-CNOT encoded'' version of $k$ linear cluster states must be another LTG state with a special biadjacency matrix sequence. 
Replacing $\gammab_{a}\to \widetilde{\Gb}_{a}$ will recover the previous argument regarding transversal measurements.
Note that CNOT operations always transform the $X$ operator to its tensor product within each $V_a$. 
The logical measurement outcome of $\mathbf{m}_{a}$ can be extracted from entries of $\widetilde{\mathbf{x}}_a$, \textit{without} performing any multiqubit gates.

\textit{Generation overhead} --
One major concern about this scheme is how to reliably prepare the all-photonic RGS (Fig.~\ref{fig:rgs}(c)). 
It could take a huge amount of resources to generate such an RGS with indeterministic methods in linear optics. 
A possible way is to use photon emitters as ancilla qubits to produce the photonic resource state \cite{Economou2010, Pichler2017, Schwartz2016, Besse2020, Wan2021PRX, Thomas2024}. 
As discussed in \cite{ButerakosPRX2017, Li2022}, if photon feedback to emitters is not allowed, at least two photon emitters are needed for generating the original unencoded RGS in \cite{Azuma2015}. 
If we apply the same approach from \cite{Li2022} to this generalized RGS, generating $\ket{\overline{\mathcal{G}}_{\Bb_{a}}}$ may maximally take $k+\mathcal{O}(1)$ emitters~\cite{SuppMat}, where the $\mathcal{O}(1)$ term is contributed by qubits of the second leaf and the tree encoding qubits. 
If one temporarily ignores the tree encoding and prepares the right arms of $\ket{\mathcal{G}_{\Bb_{a}}}$ after qubits of the left arms have been produced, the minimal emitter number is given by $k+1$. 
One can also prepare the left and right arms in an alternative order, which increases the emitter number overhead to $2k+\mathcal{O}(1)$.

\textit{Performance of QKD} --
Our new repeater scheme is particularly useful if a high rate of ebits is in demand.
We will benchmark our scheme with the model of entanglement-based QKD~\cite{E92}. 
For simplicity, we assume the only errors are photon loss during transmission and the failure of BSMs. 
In this case, Alice and Bob can immediately measure their qubits in the Pauli basis once $|\overline{\mathcal{G}}_{\Bb_{0}}\rangle$ and $|\overline{\mathcal{G}}_{\Bb_{N_R+1}}\rangle$ have arrived and process their data based on the heralded information afterward.
Particularly, given the success probability $p_s$ of a protocol, we are interested in the expected number of ebits $p_s k$ it can deliver when varying $n$.
Denote the attenuation length, the repeater interval, and the total transmission distance as $L_{\mathrm{att}}$, $L_0$ and $L$, respectively.
$L$ is evenly divided by $N_R$ repeater nodes such that $L=(N_R+1)L_0$. 
We fix $L_0 = L_{\mathrm{att}}/10$, such that $p_f = 1-\frac{1}{2}e^{-\frac{1}{10}}\approx 0.55$. 
The choice of $\bb$ typically has a complicated effect on the success probability of logical $X/Z$ measurement.
The tree encoding branch vector is set to be $\bb=(5,11,4)$ for all RGSs.
As illustrated in Fig.~\ref{fig:curves}(b), for a proper $(n,k)$, our generalized RGS with random codes (solid curves) can easily outperform the original RGS scheme (dashed curves) regarding the ratio $p_s k/n$.
Note that one can repeatedly employ the same protocol and keep those successful events. 
It is necessary to show the optimal choice of $k$ and $n$ that optimize $p_s k/n$. 
The solid curves in Fig.~\ref{fig:curves}(c) display the optimal $p_s k/n$ with the solution $n=n_*$ for given values of $k$. 
The purple shaded area (which depends on $\mathbf{b}$ and $L_0/L_{\mathrm{att}}$,\textit{ et al}.) indicates the upper bound of these curves if we further optimize the values of $k$.
It demonstrates that our new approach (solid curves) still surpasses the old approach (gray dashed curves) by a big gap over a wide range of parameters.
Fig.~\ref{fig:curves}(d,e) show the result of $p_s k$ and $p_s$ at $L=100L_{\mathrm{att}}$ under different $n$ and $k$. They display an optimal ratio $k/n$ that maximizes the outcome for a desired $k$.  
Overincreasing $n$ will deteriorate $p_s$ since the logical error of the constant-sized tree encoding becomes considerable in the case of large $n$.
The detailed calculation and comparisons are presented in~\cite{SuppMat}.

\textit{Discussion} --
In this work, we presented a new class of RGS inspired by classical error-correcting codes, optimized to suppress the erasure error due to the lossy BSM. 
The utilizable rate of BSMs per RGS is made finite for the first time even when the loss of BSM is over $50\%$. We observe significant performance improvements in the model analysis of long-range QKD.
This scheme is particularly useful if the failure of the fusion operation is the dominant error and a higher entanglement rate is desired. 
However, as the system scales up, other types of errors will become the bottleneck for entanglement distribution. 
The exploration of such resource trade-offs with improved designs will be addressed in future works.

\begin{acknowledgments}
We thank Qian Xu, Pei Zeng for their helpful discussions.
We acknowledge support from the ARO(W911NF-23-1-0077), ARO MURI (W911NF-21-1-0325), AFOSR MURI (FA9550-19-1-0399, FA9550-21-1-0209, FA9550-23-1-0338), DARPA (HR0011-24-9-0359, HR0011-24-9-0361), NSF (OMA-1936118, ERC-1941583, OMA-2137642, OSI-2326767, CCF-2312755), NTT Research, Samsung GRO, Packard Foundation (2020-71479), and the Marshall and Arlene Bennett Family Research Program. This material is based upon work supported by the U.S. Department of Energy, Office of Science, and National Quantum Information Science Research Centers.
\end{acknowledgments}

\clearpage

\widetext
\begin{center}
\textbf{\large Supplemental Material: ``Generalized Quantum Repeater Graph State''}
\end{center}
\setcounter{equation}{0}
\setcounter{figure}{0}
\setcounter{table}{0}
\setcounter{theorem}{0}
\setcounter{page}{1}
\makeatletter
\renewcommand{\theequation}{S\arabic{equation}}
\renewcommand{\thefigure}{S\arabic{figure}}
\renewcommand{\bibnumfmt}[1]{[#1]}
\renewcommand{\citenumfont}[1]{#1}

\section{The Type-II fusion operation}\label{sec:fusion}
This section demonstrates the subroutine of combining RGSs as a linear trellis graph (LTG), as displayed in the Fig.~1(d,e) of the main text.

\begin{figure}[b]
    \centering
    \includegraphics[width=0.3\linewidth]{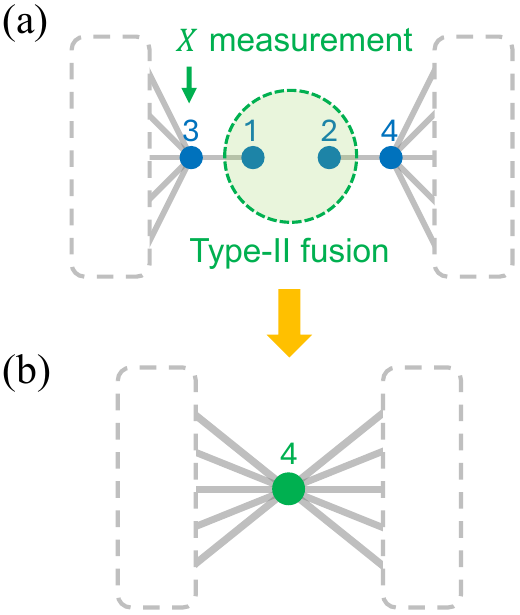}
    \caption{(a) Illustration of the fusion operation (green circle) on dangling qubits $1$ and $2$. 
    The boxes (gray dashed lines) represent arbitrary graphs.
    (b) The final desired state if a successful fusion operation, the $X$ measurement on the qubit $3$, and relevant single-qubit operations are performed. }
    \label{fig:fusion}
\end{figure}

In linear optics, the Bell state measurement can be implemented using a fusion operation~\cite{fusiongates}.
If the qubit is encoded by the polarization of photons, the type-II fusion gate can be realized with a polarizing beam splitter and polarization-discriminating photon counters.  
We are particularly interested in transforming the graph state in Fig.~\ref{fig:fusion}(a) to Fig.~\ref{fig:fusion}(c) by fusion gates and single-qubit measurements and rotations.   
The operation details are displayed in the following subsections.  

Let's start with the initial graph state in Fig.~\ref{fig:fusion}(a), and employ an ideal type-II fusion operation on dangling qubits $1$ and $2$. 
If the fusion succeeds (with probability $1-p_f$, for instance), it employs a Kraus operator
\begin{equation}
    \bra{++}_{12} + (-1)^{s_f} \bra{--}_{12}
\end{equation}
on the initial state, where $s_f\in\{0,1\}$ is determined by the fusion readout from the polarization-discriminating photon number counter. 
$\ket{\pm}:=\frac{\ket{0}+\ket{1}}{\sqrt{2}}$ is the eigen-basis of the $X$ operator.
After the successful fusion, one should further measure qubits $3$ by an $X$ measurement. 
Denote the measurement readout as $s_3\in\{0,1\}$, where $0$ and $1$ correspond to the $\ket{+}$ and $\ket{-}$ projections, respectively.
With an additional single-Pauli rotation $Z_4^{s_{3}+s_f}$ implemented in the post-measurement state, one finally obtains the desired graph state Fig.~\ref{fig:fusion}(b).

If the fusion fails at the beginning, the logical qubits $1$ and $2$ are equivalently measured in the $X$ basis. 
This failure information is heralded so logical qubits associated with vertices $3$ and $4$ are pruned by logical $Z$ measurements. 

\section{Linear trellis graphs}
\subsection{The proof of Theorem~1}
For an $n$-qubit system, the stabilizer operator
\begin{equation}
    P = \pm X^{x_1}Z^{z_1}\otimes X^{x_2}Z^{z_2}\otimes\cdots
    \otimes X^{x_n}Z^{z_n}
\end{equation}
is a Hermitian element in the Pauli group,
where $x_i,z_i\in\mathbb{F}_2$ can be gathered as a binary vector in $\mathbb{F}_2^{2n}$:
\begin{equation}
    [x_1,x_2,\cdots,x_n|z_1,z_2,\cdots,z_n]\;,
\end{equation}
if the global phase of $P$ is ignored or stored separately.
Since a pure stabilizer state has $n$ independent stabilizer operators, their corresponding binary vectors are collected as a matrix in $\mathbb{F}_2^{n\times 2n}$, which is called the tableau matrix.
For instance, the following three stabilizer operators
\begin{equation}
\begin{aligned}
    X\otimes Z \otimes I \equiv \hat{X}_1 \hat{Z}_2 \;,\\
    Z\otimes X \otimes Z \equiv \hat{Z}_1 \hat{X}_2 \hat{Z}_3 \;,\\
    I\otimes Z \otimes X \equiv \hat{Z}_2 \hat{X}_3
\end{aligned}
\end{equation}
can be represented by the following $3\times 6$ tableau matrix:
\begin{equation}
\left[
\begin{array}{ccc|ccc}
    1&0&0&0&1&0\\
    0&1&0&1&0&1\\
    0&0&1&0&1&0\\
\end{array}
\right].\;
\end{equation}
One can simply call the left and right half of the tableau matrix as $X$-block and $Z$-block.
For a qubit graph state $\ket{\mathcal{G}}$ of a graph $\mathcal{G}=(V,E)$, there are $|V|$ stabilizers $\hat{K}_i$:
\begin{equation}
    \hat{K}_i := \hat{X}_i\prod_{j\in\mathcal{N}(i)}\hat{Z}_j\;.
\end{equation}
Let $\boldsymbol{\Gamma}$ be the adjacency matrix of $\mathcal{G}$, that is, $\boldsymbol{\Gamma}_{ij}=1$ if and only if $(i,j)\in E$, otherwise, $\boldsymbol{\Gamma}_{ij}=0$. 
Denote $\mathbf{I}_k$ as the $k\times k$ identity matrix and $\mathbf{0}_k$ as the $k\times k$ zero matrix.
It follows that the corresponding tableau matrix for $\ket{\mathcal{G}}$ is written as:
\begin{equation}
    \mathbf{T}_{\mathcal{G}} = \left[\;\mathbf{I}_{|V|}\;|\;\boldsymbol{\Gamma}\;\right]\;.
\end{equation}

Any Clifford operation on the stabilizer state can be represented by a linear transformation of the tableau matrix \cite{Aaronson2004PRA}. 
However, a fundamental row transformation on the tableau matrix does not change the stabilizer state, as the stabilizer group is unchanged.
Suppose in the statement of Theorem~1, there are $\ell$ biadjacency matrices $\betab_a$ and $\ell + 1$ vertex subsets $V_a$. 
Denote $n_a:=|V_a|$, the theorem regarding the linear trellis graphs (LTG) and the proof are demonstrated as follows.

\begin{theorem}
    Let $\mathcal{G}_{\mathrm{tr}}$ be an LTG associated with $\{V_a\}$ and $\betab_{a} = \gammab_{a}^T\gammab_{a+1}$, where all $\gammab_{a}\in\mathbb{F}_2^{k\times n_a}$ have full row rank. Then the graph state $\ket{\mathcal{G}_{\mathrm{tr}}}$ is local-CNOT equivalent to another graph state $\ket{\mathcal{G}'_{\mathrm{tr}}}$ of an LTG $\mathcal{G}'_{\mathrm{tr}}$ with $\{V_a\}$ and $\betab'_{a} = \gammab'^{T}_{a} \gammab'_{a+1}$, where $\gammab'_{a}=[\mathbf{I}_k,\mathbf{0}_{k,n_a-k}]$ and $\mathbf{0}_{k,m}$ is a $k\times m$ zero block.
\end{theorem}

\begin{proof}
     Let the $i$th row of $\mathbf{T}_{\mathcal{G}_{\text{tr}}}$ represent the $i$th stabilizer $\hat{K}_i := \hat{X}_i\prod_{j\in\mathcal{N}(i)}\hat{Z}_j$.
     The tableau matrix of the graph state $\ket{\mathcal{G}_{\text{tr}}}$ is
     \begin{equation}\label{eq:tableau_G_0}
        \arraycolsep=1.2pt \def\arraystretch{0.9}
        \mathbf{T}_{\mathcal{G}_{\text{tr}}}=\left[
        \;\;\;\mathbf{I}_{|V|}\;\;\;
        \left|
        \begin{array}{ccccccc}
            \mathbf{0}_{n_1}&\betab_1&&&&\\
            \betab_1^T&\mathbf{0}_{n_2}&\betab_2&&&\\
            &\ddots&\ddots&\ddots&\\
            &&\betab_{\ell-1}^T&\mathbf{0}_{n_{\ell}}&\betab_{\ell}\\
            &&&\betab_{\ell}^T&\mathbf{0}_{n_{\ell+1}}\\
        \end{array}
        \right.
        \right].
    \end{equation}
    One can ignore the global phase of each stabilizer operator, since (i) CNOT operations do not change the phase, (ii) the row operation used in this proof does not multiply different Pauli operators that yield extra phase.

    Note that $\betab_a = \gammab_a^T \gammab_{a+1}$ ($1\le a \le \ell$) and that $\gammab_a$ has full row rank. This implies there exists a set of invertible matrices $\{\Qb_a: \Qb_{a}\in\mathbb{F}_2^{n_a\times n_a},\;1\le a \le \ell+1\}$ such that 
    \begin{equation}\label{eq:gammaQ}
        \gammab_a' :=
        \gammab_a \Qb_{a} = 
        \begin{bmatrix}
            \mathbf{I}_{k}&\mathbf{0}_{k,n_a-k}
        \end{bmatrix}\;,
    \end{equation}
    for all $a$, which means
    \begin{equation}
        \Qb_{a}^T\betab_a \Qb_{a+1} = 
        \begin{bmatrix}
            \mathbf{I}_{k}&\\
            &\mathbf{0}_{n_a-k,n_{a+1}-k}
        \end{bmatrix}\;.
    \end{equation}
    Denote
    \begin{equation}
      \begin{aligned}
            \betab'_a &:= \Qb_{a}^T\betab_a \Qb_{a+1} = \gammab_a'^{T}\gammab_{a+1}'\;,
            \\
            \Qb &:= \mathrm{diag}(\Qb_1,\cdots, \Qb_{\ell+1})\;,
            \\
            \mathbf{C} &:= \mathrm{diag}\big[(\Qb^{-1})^T, \Qb\big]\;,
      \end{aligned}
    \end{equation}
    then $\mathbf{T}_G$ in Eq.~\eqref{eq:tableau_G_0} can be transformed as:
    \begin{equation}\label{eq:QTC}
        \arraycolsep=1.2pt \def\arraystretch{1.0} 
        \Qb^T\mathbf{T}_{\mathcal{G}_{\text{tr}}}\mathbf{C}
        =
        \left[
        \;\;\; \mathbf{I}_{|V|} \;\;\;
        \left|
        \begin{array}{cccccc}
            \mathbf{0}_{n_1}&\betab'_1&&&\\
            \betab_1'^{T}&\mathbf{0}_{n_2}&\betab'_2&&&\\
            &\ddots&\ddots&\ddots&&\\
            &&\betab_{\ell-1}'^{T}&\mathbf{0}_{n_\ell}&\betab'_{\ell}&\\
            &&&\betab_{\ell}'^{T}&\mathbf{0}_{n_{\ell+1}}\\
        \end{array}
        \right.
        \right],
    \end{equation}
    where the column transformation $\mathbf{C}$ represents the local-CNOT operation within each $V_a$. (When the $Z$-block is column-transformed by $\Qb_a$, the corresponding $X$-block is column-transformed by $(\Qb_a^{-1})^T$~\cite{Aaronson2004PRA}.) 
    Therefore, it proves the sufficiency.
      
    Conversely, starting from a tableau matrix of the r.h.s. of Eq.~\eqref{eq:QTC} with $\betab'_{a} = \gammab'^{T}_{a} \gammab'_{a+1}$, and $\gammab'_{a}=[\mathbf{I}_k,\mathbf{0}_{k,n_a-k}]$. 
    Any local-CNOT operations can convert it back to the form of Eq.~\eqref{eq:tableau_G_0} with $\betab_a = \gammab_a^T \gammab_{a+1}$.
    Hence, it proves the necessity.
\end{proof}

In the above proof, since all $1$'s in $\betab'_a$ and $\betab'_{a+1}$ are aligned,
the stabilizers given by $\Qb^T\mathbf{T}_{\mathcal{G}_{\text{tr}}}\mathbf{C}$ describe a graph state that contains $k$ parallel linear cluster states and some decoupled $\ket{+}$ qubits.
    
\subsection{An example of local-CNOT equivalence}
\begin{figure}[t]
    \centering
    \includegraphics[width=0.5\linewidth]{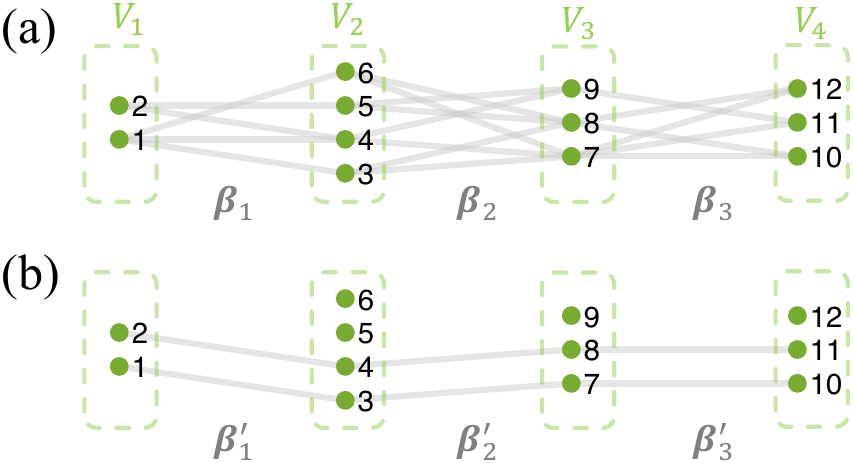}
    \caption{This figure illustrates the local-CNOT equivalence associated with $\{V_a\}$ between two linear trellis graphs (a) and (b), which are separately defined by biadjacency matrix sets $\{\betab_a\}$ and $\{\betab'_a\}$.
    The boxes with dashed lines divide the vertices into different $V_a$'s.}
    \label{fig:LCE}
\end{figure}
We demonstrate the concept of local-CNOT equivalence between two LTGs in Fig.~\ref{fig:LCE}. 
In the small example in Fig.~\ref{fig:LCE}(a), there are $|V_1|=2$, $|V_2|=4$, and $|V_3|=|V_4| = 3$. 
The biadjacency matrices are
\begin{equation}
    \betab_1 = \gammab_1^T\gammab_2,\quad
    \betab_2 = \gammab_2^T\gammab_3,\quad
    \betab_3 = \gammab_3^T\gammab_4,
\end{equation}
where
\begin{equation}
    \begin{aligned}
        \gammab_1&=\begin{bmatrix}
            1&0\\
            0&1\\
        \end{bmatrix}\;, \quad
        \gammab_2=\begin{bmatrix}
            1&1&0&1\\
            0&1&1&0\\
        \end{bmatrix}\;, \\
        \gammab_3&=\begin{bmatrix}
            1&1&0\\
            0&1&1\\
        \end{bmatrix}\;, \quad
        \gammab_4=\begin{bmatrix}
            1&1&1\\
            0&1&0\\
        \end{bmatrix}\;. \\
    \end{aligned}
\end{equation}
Thus, the $\Qb_a$ matrix satisfying Eq.~\eqref{eq:gammaQ} is found below:
\begin{equation}
    \begin{aligned}
        \Qb_1&=\begin{bmatrix}
            1&0\\
            0&1\\
        \end{bmatrix}\;, \quad
        \Qb_2=\begin{bmatrix}
            1&1&1&0\\
            0&1&1&1\\
            0&0&1&1\\
            0&0&0&1\\
        \end{bmatrix}\;, \\
        \Qb_3&=\begin{bmatrix}
            0&0&1\\
            1&0&1\\
            1&1&1
        \end{bmatrix}\;, \quad
        \Qb_4=\begin{bmatrix}
            1&0&1\\
            0&1&0\\
            0&1&1
        \end{bmatrix}\;.
    \end{aligned}
\end{equation}
One can decompose $\Qb_a$ by elementary XOR operations on columns. 
For example:
\begin{equation}
    \Qb_2
    =
    \begin{bmatrix}
            1&&&\\
            &1&&\\
            &&1&\\
            1&&&1\\
    \end{bmatrix}\!\!\!
    \begin{bmatrix}
            1&&&\\
            1&1&&\\
            &&1&\\
            &&&1\\
    \end{bmatrix}\!\!\!
    \begin{bmatrix}
            1&&&\\
            &1&&\\
            &1&1&\\
            &&&1\\
    \end{bmatrix}\!\!\!
    \begin{bmatrix}
            1&&&\\
            &1&&\\
            &&1&\\
            &&1&1\\
    \end{bmatrix}
\end{equation}
corresponds to the following controlled-NOT operations within $V_2$:
\begin{equation}
\mathrm{CNOT}_{34}\cdot
\mathrm{CNOT}_{23}\cdot
\mathrm{CNOT}_{12}\cdot
\mathrm{CNOT}_{14}\;.
\end{equation}
Similarly, $\Qb_1$ corresponds to the identity operation within $V_1$.
$\Qb_3$ corresponds to the following operation within $V_3$:
\begin{equation}
\mathrm{CNOT}_{31}\cdot
\mathrm{CNOT}_{13}\cdot
\mathrm{CNOT}_{32}\cdot
\mathrm{CNOT}_{23}\;.
\end{equation}
$\Qb_4$ corresponds to the following operation within $V_4$:
\begin{equation}
\mathrm{CNOT}_{32}\cdot
\mathrm{CNOT}_{13}\cdot
\mathrm{CNOT}_{12}\;.
\end{equation}
After the above local-CNOT operation, there are 
\begin{equation}
    \begin{aligned}
        \betab_1'&=\begin{bmatrix}
            1&0&0&0\\
            0&1&0&0\\
        \end{bmatrix},\;\;
        \betab_2'=\begin{bmatrix}
            1&0&0\\
            0&1&0\\
            0&0&0\\
            0&0&0\\
        \end{bmatrix},\;\;
        \betab_3'&=\begin{bmatrix}
            1&0&0\\
            0&1&0\\
            0&0&0
        \end{bmatrix},
    \end{aligned}
\end{equation}
which describe the LTG in Fig.~\ref{fig:LCE}(b).

\section{Derivation of $\avg{\epsd}$}
We will first consider the value of $\avg{\epsd}$ for matrices $\widetilde{\Gb}$ sampled uniformly at random from $\mathbb{F}_2^{k\times \widetilde{n}}$.
The probability that $\mathrm{rank}_2(\widetilde{\Gb}) < k$ is well known as
\begin{equation}
    \epsd(k,\widetilde{n})
    :=
    1-\prod_{i=0}^{k-1}(1-2^{i-\widetilde{n}})\quad \forall \widetilde{n}\in\mathbb{N}\;.
\end{equation}
It can be derived by randomly sampling $k$ vectors uniformly from $\mathbb{F}_2^{\widetilde{n}}$. 
The probability that the $(r+1)$-th row $(r\ge1)$ is independent of the previous $r$ rows is $1-2^{(r-1)-\widetilde{n}}$.
Multiplying the probabilities of these independent events yields the probability of $\mathrm{rank}_2(\widetilde{\Gb}) = k$.
When a general erasure code is used \cite{RaptorCodes,Luby2001EECC}, $\epsd(k,\widetilde{n})$ may not have a compact expression.

Secondly, we consider another random matrix $\Gb$ uniformly sampled from $\mathbb{F}_2^{k\times n}$,
and $\widetilde{\Gb}$ is obtained from $\Gb$ by randomly removing each column with erasure rate $p_f$.
Then $\widetilde{\Gb}$ is equivalently a matrix uniformly sampled from $\mathbb{F}_2^{k\times \widetilde{n}}$, where the column number $\widetilde{n}\sim B(n,(1-p_f))$ follows the binomial distribution.
Therefore, the rank-deficient probability $\avg{\epsd}$ in Fig.~2(a) is given by the expectation of $f(k,\widetilde{n})$:
\begin{equation}\label{eq:prob_DeficientRank}
    \begin{aligned}
    \avg{\epsd} &\equiv \mathrm{Pr}(\mathrm{rank}_2(\widetilde{\Gb})<k|\Gb\in\mathbb{F}_2^{k\times n})
    =\sum_{\widetilde{n}=0}^{n}
    \epsd(k,\widetilde{n})f_{\mathrm{bin}}(\widetilde{n};n,1-p_f)\;,
    \end{aligned}
\end{equation}
where
\begin{equation}\label{eq:fbin}
    f_{\mathrm{bin}}(\widetilde{n};n,q):=\binom{n}{\widetilde{n}}q^{\widetilde{n}}(1-q)^{n-\widetilde{n}}
\end{equation}
is the probability mass function of the binomial distribution.
The above $\epsd$ transits from $0$ to $1$ within an asymptotically vanishing region around $k/n = 1-p_f$.
This is due to $\epsd(k,\widetilde{n})$ dropping from $1$ to $0$ \textit{exponentially} with respect to $\widetilde{n}-k$, whose decay length is independent of $k$ and $\widetilde{n}$.
On the other hand, the binomial distribution (Eq.~\eqref{eq:fbin}) has a width (the standard deviation) $\sigma_{\widetilde{n}}=\mathcal{O}(\sqrt{n})$ for $\widetilde{n}$ around the peak $(1-p_f)n$.
Therefore, the inner product of the two --- Eq.~\eqref{eq:prob_DeficientRank} has a transitional region for $k/n$ near $1-p_f$, with width $\sigma_{k/n}=\mathcal{O}(1/\sqrt{n})=\mathcal{O}(1/\sqrt{k})$, which is asymptotically vanishing if $p_f>0$ and $k\to \infty$.

\section{Bipartition entropies of RGS}
Ref.~\cite{Li2022} has proved the minimal number of emitter resources needed to generate a photonic graph state deterministically. 
Specifically, if emitter qubits generate each photonic qubit by a Kraus operator
\begin{equation}\label{eq:emissionkraus}
    \ket{00}_{ep}\!\!\bra{0}_e + \ket{11}_{ep}\!\!\bra{1}_e = \mathrm{CNOT}_{ep}\ket{0}_p
\end{equation}
at the emission stage ($e$ represents the emitter qubit and $p$ represents the photonic qubit), and all operations are Clifford, then the minimal emitter number required is given by the maximum of the height function $h(x)$.
Assuming there are $N$ qubits in the final photonic state, the height function $h(x)$ is defined by the bipartition entropy of the final graph state's qubit sets $[1,x]$ and $[x+1,N]$.
Therefore, one can optimize the emission ordering to reduce the emitter number overhead $h_{\max}$.

\begin{figure}[t]
    \centering
    \includegraphics[width=0.5\linewidth]{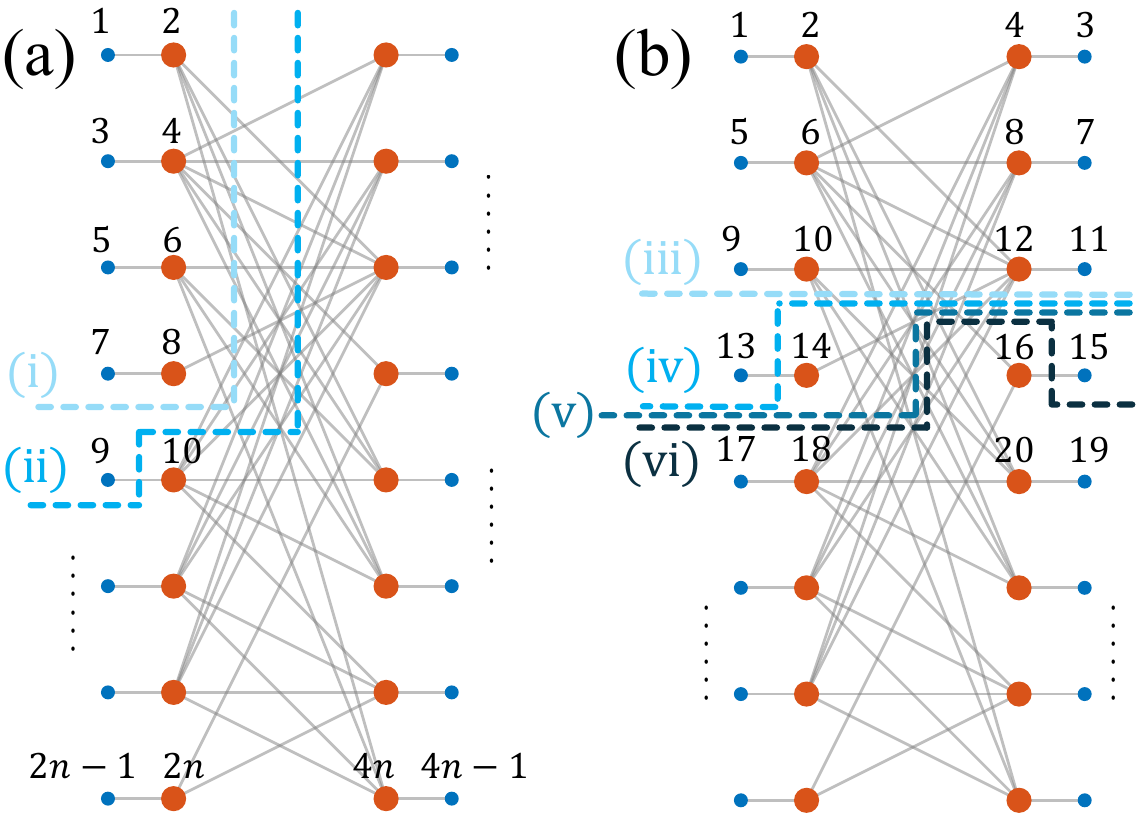}
    \caption{Different ways of partitioning (dashed lines) the generalized RGS into two sets of qubits.}
    \label{fig:bipart}
\end{figure}

Fig.~\ref{fig:bipart} demonstrates two possible emission orderings, which are not only symmetric for possible practical usage but also have reasonably small maxima of $h(x)$.
In Fig.~\ref{fig:bipart}(a), if the rank of the biadjacency matrix (cut-rank) between the red vertices is $k$, then the partition line (i) at most provides a cut-rank $k$, while the partition line (ii) at most provides a cut-rank $k+1$ due to the additional second leaf. 
Since the red node is encoded by tree graphs, the partition line that cuts through the tree yields a higher cut-rank, which is usually proportional to the depth of the tree. 
If the tree-encoding is fixed for all qubits of the 1st leaf, then we have $h_{\max}=k+\mathcal{O}(1)$.

For the ordering in Fig.~\ref{fig:bipart}(b), we first focus on the partition lines (iii) and (v).
These lines divide the entire RGS into the upper (U) and lower (D) parts. 
Together with the division of the left (L) and right (R) arms, the RGS is divided into 4 parts, labeled as: upper-left (UL), lower-left (DL), upper-right (UR), and lower-right (DR).
Therefore, we can denote the original rank-$k$ biadjacency matrix $\Bb$ as:
\begin{equation}\label{eq:Bquarter}
    \Bb=
    \begin{bmatrix}
        \Bb_{(\text{UL},\text{UR})} & \Bb_{(\text{UL},\text{DR})} \\
        \Bb_{(\text{DL},\text{UR})} & \Bb_{(\text{DL},\text{DR})} \\
    \end{bmatrix}\;,
\end{equation}
in which $\Bb_{(\mathcal{S}_1,\mathcal{S}_2)}$ is the biadjacency matrix between vertex subsets $\mathcal{S}_1$ and $\mathcal{S}_2$. 
For example, $\Bb_{(\text{LD},\text{RU})}$ is the biadjacency matrix between the lower-left part and the upper-right part.
Note that each submatrix $\Bb_{(\mathcal{S}_1,\mathcal{S}_2)}$ in Eq.~\eqref{eq:Bquarter} possibly has a rank equal to $k$, if its size is sufficiently large.
The biadjacency matrix between $\text{UL}\cup \text{UR}$ and $\text{DL}\cup \text{DR}$ is given by
\begin{equation}
    \begin{bmatrix}
        \Bb_{(\text{UL},\text{DR})} &  \\
         & \Bb^T_{(\text{DL},\text{UR})} \\
    \end{bmatrix}\;,
\end{equation}
which at most has a rank equal to $2k$.
Therefore, partition lines (iv) and (vi) at most have a cut-rank equal to $2k+1$ due to the additional qubit on the second leaf.
Similarly, if each qubits on the 1st leaf are encoded by the same fixed tree graph, we have $h_{\max}=2k+\mathcal{O}(1)$.

Finally, the above discussion can also be generalized so it applies to the case where the number of left and right arms are imbalanced, such as the boundary RGS $|\overline{\mathcal{G}}_{\Bb_0}\rangle$ and $|\overline{\mathcal{G}}_{\Bb_{N_R}}\rangle$.

\section{The QKD efficiency}\label{sec:QKDsim}
\begin{figure}[t]
    \centering
    \includegraphics[width=0.9\linewidth]{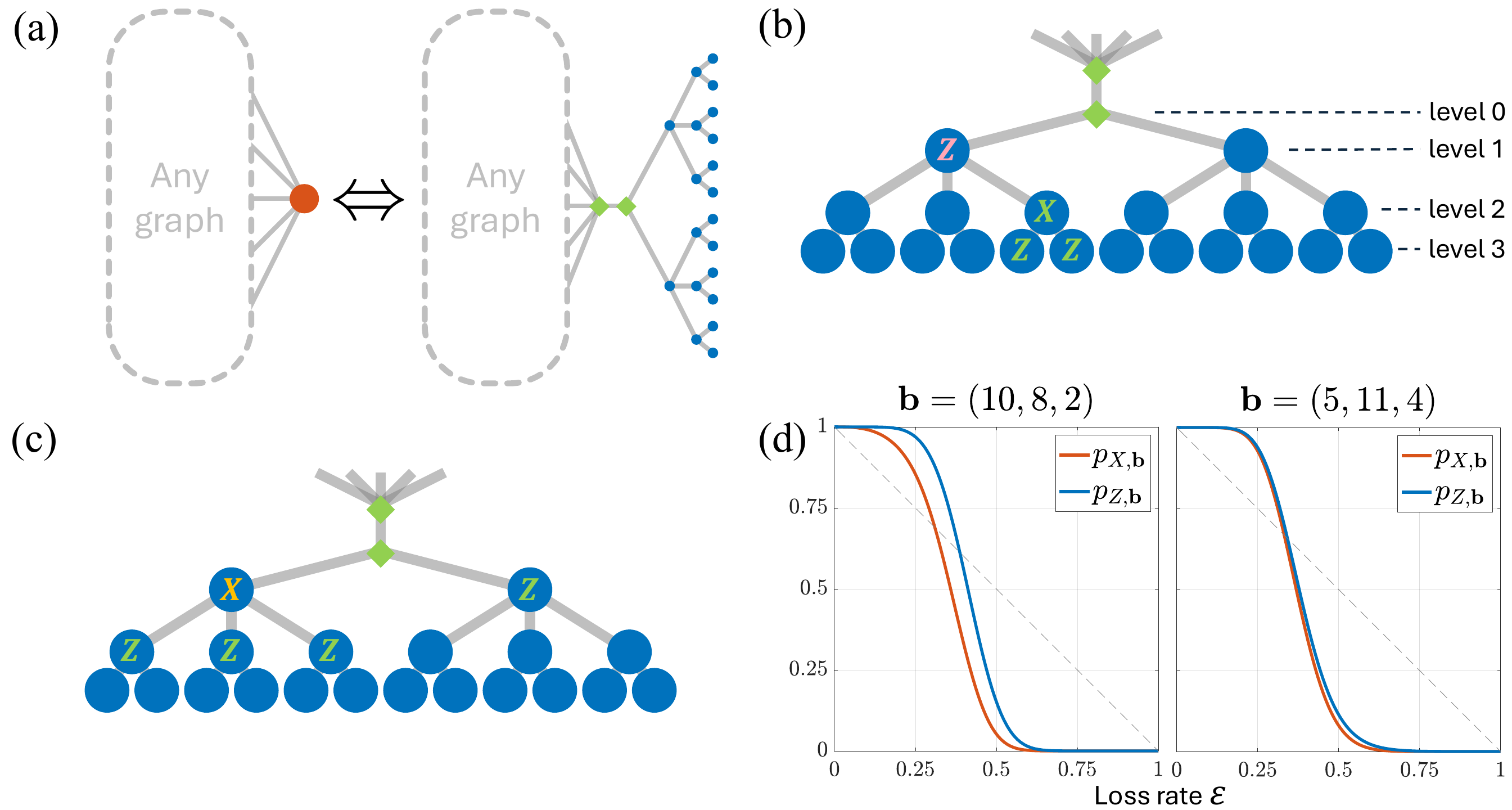}
    \caption{This figure shows how the logical $X$ and $Z$ measurements are performed based on \cite{Varnava2006}. 
    Qubits labeled by green diamond nodes are pre-measured by $X$ measurements. 
    All blue vertices represent physical qubits.
    Panels (a), (b), and (c) use the same tree of the branch vector $\bb = (2,3,2)$.
    (a) An example of tree graph encoding, where the red vertex (left) represents a tree graph attached to the same vertex (right). 
    (b) The logical $Z$ measurement is given by \textit{all} successful direct or indirect $Z$ measurements at level $1$.
    An example of the indirect $Z$ measurement (pink letter) at level $1$ is performed by the $X$ and $Z$ measurements (green letters) on the `leaf qubits' at the following two levels. Note that any $Z$ measurements at lower levels can also be indirect $Z$ measurements defined similarly.
    (c) The logical $X$ measurement on the ``root qubit'' at level $1$ is given by any successful direct $X$ measurement (yellow letter) at level $1$, together with other direct or indirect $Z$ measurements that prune the rest of the graph.
    (d) Two examples of $p_{X,\bb}$ and $p_{Z,\bb}$ versus the loss rate $\varepsilon$, obtained from different branch vectors.}
    \label{fig:tree}
\end{figure}
To compare the actual performance of our new scheme, we consider using photonic repeater graph states for entanglement-based QKD. 
The dominant error to be corrected is the erasure error, which includes the failure to receive a photon and perform a BSM. 
Let's neglect any Pauli error, and assume that the probability of receiving a photon sent over distance $x$ is $e^{-x/L_{\mathrm{att}}}\equiv 1 - \varepsilon$, where $L_{\mathrm{att}}$ is the attenuation length, and $\varepsilon$ is the photon loss rate. 

Fig.~\ref{fig:tree}(a) displays an example of a tree encoding with $\bb = (2,3,2)$.
If the photonic qubit is encoded in tree graphs~\cite{Varnava2006} with the branch vector $\bb := (b_0,b_1,b_2,\cdots,b_m)$, the success rate of performing a logical $X$ and $Z$ measurements are given by $p_{X,\bb}(\varepsilon)$ and $p_{Z,\bb}(\varepsilon)$ in terms of the qubit loss rate $\varepsilon$:
\begin{equation}\label{eq:pxpzR}
    \begin{aligned}
    p_{Z,\bb}(\varepsilon) &= (1-\varepsilon+\varepsilon R_1)^{b_0},\\
    p_{X,\bb}(\varepsilon) &= [(1-\varepsilon+\varepsilon R_1)^{b_0} - (\varepsilon R_1)^{b_0}](1-\varepsilon + \varepsilon R_2)^{b_1},\\
    R_\ell = 1 &- \left[1-(1-\varepsilon)(1-\varepsilon+\varepsilon R_{\ell+2})^{b_{\ell+1}}\right]^{b_{\ell}},\\
    \end{aligned}
\end{equation}
where $R_\ell$ $(0\le \ell \le m)$ can be solve iteratively, with $R_{\ell}=0$ for $\ell\ge m+1$, $b_{m+1}=0$. 
In Eq.~\eqref{eq:pxpzR}, $R_\ell$ is the success probability of indirectly measuring any given qubit at level $\ell$ in the $Z$ basis.
The indirect $Z$ measurements are demonstrated in Fig.~\ref{fig:tree}(b). 
The expression of $p_{Z,\bb}$ can be understood as the power of $b_0$ of all direct or indirect successful $Z$ measurements at level $1$, where the successful $Z$ measurement on any qubit at level $1$ is simply $(1-\varepsilon + \varepsilon R_1)$.
The derivation of $p_{X,\bb}$ can be found in \cite{Varnava2006}.

In the RGS protocol, suppose we are using a random matrix $\Gb_a$ for the erasure code ($\Gb_a$ is uniformly sampled from $\mathbb{F}_2^{k\times n}$).
The interval of repeater stations is $L_0$.
Then the success rate of a repeater station obtaining a full-rank $\widetilde{\Gb}_a$ and then performing all required logical measurements is given by
\begin{equation}
    \begin{aligned}
    &r_{\mathrm{RRGS}}(L_0,k,n,\bb):=\sum_{\widetilde{n}=0}^n[1-\epsd(k,\widetilde{n})]
    f_{\mathrm{bin}}(\widetilde{n};n,1-p_f) p_{X,\bb}(\varepsilon)^{2\widetilde{n}}
    p_{Z,\bb}(\varepsilon)^{2(n-\widetilde{n})}\;,
    \end{aligned}
\end{equation}
where $p_f := 1-\frac{1}{2}e^{-L_0/L_{\mathrm{att}}}\ge 0.5$ is the failure rate of a fusion operation since the photon loss during transmission also contributes to the failure of the fusion operation. 
The physical qubit loss rate here is $\varepsilon = 1-e^{-L_0/(2L_{\mathrm{att}})}$.
The factor $p_{X,\bb}^{2\widetilde{n}}$ originates from post-selecting the experimental outcome with the success of all logical $X$ measurements. 
We indicated that this strategy is sub-optimal since the final $k$ Bell states are extracted as long as all $\mathbf{m}_a$'s are decoded, which does not require all logical $X$ measurements to succeed. 
A better strategy that has this additional fault tolerance comes with a $\Gb_a$ which has more redundant columns.
Therefore, the obtained $\widetilde{\mathbf{x}}_a$ has more bits when it is slightly corrupted (bit flip or bit loss). A well-conceived decoder can ensure $\mathbf{m}_a$ being learned by Alice and (or) Bob from the corrupted $\widetilde{\mathbf{x}}_a$.
In this work, we keep the calculation simple by ignoring these better options, as the result already displays evidence of outperformance.

If $N_{\mathrm{R}}$ repeater stations are used for a total distance $L$, then $N_{\mathrm{R}}+1$ resource states are distributed.
We assume each photon travels from the middle of two repeater stations. 
As for the two ends of Alice and Bob, we assume they each measure the received $k$ logical qubits in the $X$ and $Z$ basis with equal probabilities. 
The joint success probability of the entire process is 
\begin{equation}\label{eq:FRGS_QKD_succ}
\begin{aligned}
    &(p_s)_{\mathrm{RRGS}}:=\left[\frac{p_{X,\bb}(\varepsilon)+p_{Z,\bb}(\varepsilon)}{2}\right]^{2k} r_{\mathrm{RRGS}}\left(L_0,k,n,\bb\right)^{N_{\mathrm{R}}}.
\end{aligned}
\end{equation}

As a comparison, the result of \cite{Azuma2015} is given as follows. In this case, using RGS $|\overline{\mathcal{G}}^n_c\rangle$, the probability of a repeater station performing at least one successful BSM and successfully carrying out all its logical measurements is
\begin{equation}
\begin{aligned}
    &r_{\mathrm{CRGS}}(L_0,n,\bb):= 
    (1-p_f^n) p_{X,\bb}(\varepsilon)^{2}
    p_{Z,\bb}(\varepsilon)^{2(n-1)}\;.
\end{aligned}
\end{equation}
The first factor represents the chance of implementing at least one BSM. The following factors arise from performing one pair of logical $X$ measurements for one of the successful BSMs and pruning the other qubits with logical $Z$ measurements.
Since the complete graph RGS encodes only one ebit, let's pre-prune the first and the last RGS so Alice and Bob only need to take care of single qubit measurements. As a result, the joint success probability of the entire process is 
\begin{equation}\label{eq:CRGS_QKD_succ}
\begin{aligned}
    (p_s)_{\mathrm{CRGS}}
    &=\left[\frac{p_{X,\bb}(\varepsilon)+p_{Z,\bb}(\varepsilon)}{2}\right]^{2} r_{\mathrm{CRGS}}\left(L_0,n,\bb\right)^{N_{\mathrm{R}}}\;.
\end{aligned}
\end{equation}
Since different protocols have different numbers of logical $X$ and $Z$ measurements, we should adopt a branch vector $\bb$ which gives $p_{X,\bb}\approx p_{Z,\bb}$ for fairness. The branch vector can be chosen to be $\bb = (5,11,4)$ to meet this approximation (see Fig.~\ref{fig:tree}(d)).

\begin{figure}[t]
    \centering
    \includegraphics[width=1.00\linewidth]{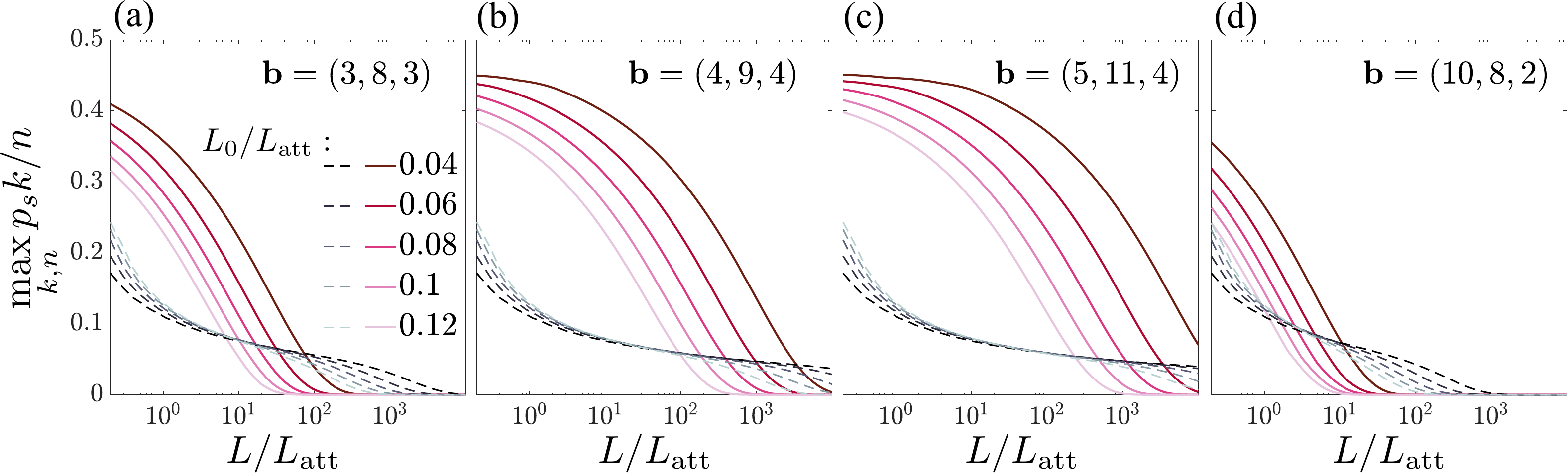}
    \caption{We present the values of $\max_{n}p_sk/n$ for the original RGS protocol (dashed curves) and the values of $\max_{k,n}p_sk/n$ for the generalized RGS protocol (solid curves). Different values of $L_0/L_{\mathrm{att}}$ are represented by different colors. Note that the results of random codes do not converge to the original RGS as $k = 1$, since the original RGS utilizes a deterministic generator matrix.}
    \label{fig:envelope}
\end{figure}

So which protocol has a better efficiency? Assuming we are in the situation with $r_{\mathrm{RRGS}}\approx r_{\mathrm{CRGS}}$, the factor $[(p_{X,\bb}+p_{Z,\bb})/2]^{2k}\sim\mathcal{O}(e^{-ck})$ in Eq.~\eqref{eq:FRGS_QKD_succ} is exponentially suppressed by $k$ with some parameter $c$.
Intuitively, using the original protocol from \cite{Azuma2015} has a better $p_s$ since $k$ is smaller. 
However, we highlight that the generalized scheme established multiple ebits at the same time, so the mean value of ebits $p_s k = \mathcal{O}(ke^{-ck})$ can be significantly better than $1$ at some $k$.
Due to the complexity of function $p_s$, we should directly compare the protocol performance.
In the main text, we have shown a better performance regarding $\max_n p_s k/n$ and its upper bound $\max_{k,n} p_s k/n$ in Fig.2(b). 
Here, we proceed to provide more instances of the upper bound $\max_{k,n} p_s k/n$ with respect to different $\bb$ in Fig.~\ref{fig:envelope}. 
We can see that our generalized RGS (solid curves) significantly outperforms the original RGS (dashed line) before some large $L$. Such outperformance is mainly affected by $\bb$. It suggests that better RGS designs may require more efficient, loss-tolerant encoding to protect against qubit loss on the ``leaves.''

\section{Fault tolerance of $\mathbf{m}_a$ and the choice of $\Gb_a$}\label{sec:ft_meas}
In our protocol, the repeater station $a$ will inform Alice and Bob of its measurement result regarding $\widetilde{\mathbf{x}}_a$. 
In our calculation, we ideally assume $\widetilde{\mathbf{x}}_a$ is uncorrupted, then solving $\mathbf{m}_a$ from $\widetilde{\mathbf{x}}_a = \mathbf{m}_a\widetilde{\Gb}_a$ is possible if $\widetilde{\Gb}_a$ is full rank.

Practically, all measurements in the repeater station should be considered erroneous. When pruning the RGS with logical $Z$ measurements, its logical error (bit flip or bit erasure) will propagate from the qubit of the $1$st leaf to its $\mathcal{O}(n)$ neighboring qubits due to the connectivity of $\Bb_a$. 
This error is likely detrimental since these neighboring qubits need to be rotated by $Z$ gates conditioned on the logical readout, resulting in a non-local error. This deficiency can be overcome by improving the tree encoding with better encoding.

However, a logical error occurs on the logical $X$ measurement will stay local. In Fig.~\ref{fig:fusion}, if logical $X$ measurements on the qubit of a pair of vertex $3$ and $4$ are erroneous (bit flip or bit erasure), they at most affect one entry of the vector $\widetilde{\mathbf{x}}_a$.  
As Alice and Bob only concern with the final $\mathbf{m}_a$, a proper choice of $\Gb_a$ can offer sufficient redundancy for the inference of $\mathbf{m}_a$ from equation $\widetilde{\mathbf{x}}_a=\mathbf{m}_a\widetilde{\Gb}_a$.
Any classical linear binary code equipped with a decoder can be a good candidate.
It implies that the logical error of the logical $X$ measurement can still be benign if it occurs at a reasonably small rate.

\begin{figure}[t]
    \centering
    \includegraphics[width=0.6\linewidth]{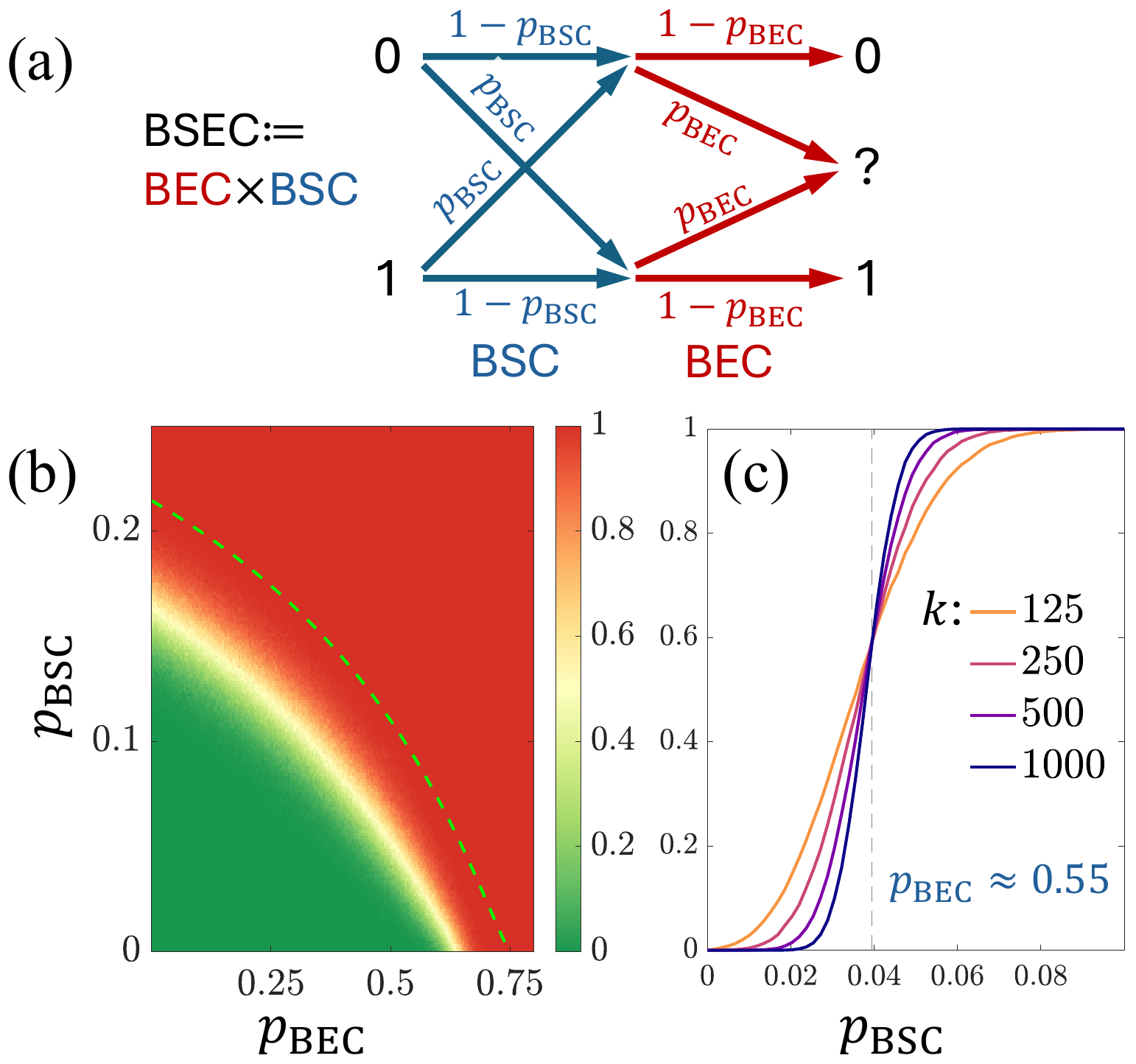}
    \caption{Panel (a) demonstrates the information corruption by BSEC. The input bit $(\mathbf{x}_a)_i=0$ or $1$ will either be received correctly or incorrectly, or lost (label as ``?''), given by certain probabilities. (b) The logical error rate or decoding failure rate (represented by the colorbar) of a rate-$1/4$ regular LDPC code, with $[n,k]=[500,125]$. The green dashed line represents the Shannon limit, above which $\mathcal{C}_{\text{BSEC}}<1/4$ (see Eq.~\eqref{eq:C_BSEC}). (c) A vertical cut of panel (b) at $p_{\mathrm{BEC}}\approx 0.55$ under various $k$ ($k/n=1/4$), the vertical axis represents the logical error rate of recovering $\mathbf{m}_a$.}
    \label{fig:ldpc}
\end{figure}

To deal with the faulty logical $X$ measurement, instead of using a uniformly sampled random code, one can use a regular low-density parity-check (LDPC) code (Gallager code) to design RGS.
The merit is that $\widetilde{\mathbf{x}}_a$ is protected from the additional corruption of both bit flip and bit erasure. 
Given a parity-check matrix $\mathbf{H}_{\mathrm{LDPC}}$ of an LDPC code, one can find its generator matrix $\Gb_{\mathrm{LDPC}}$. 
Specifically, we simply can set $\mathcal{\Gb}_a = \Gb_{\mathrm{LDPC}}$ for $1\le a \le N_R$. 
We model the noise channel with both bit flip and bit erasure using the binary symmetric erasure model (BSEC), which is equivalently the concatenation of a binary symmetric channel (BSC) with bit flip rate $p_{\mathrm{BSC}}$, and a binary erasure channel (BEC) with bit erasure rate $p_{\mathrm{BEC}}$ (see Fig.~\ref{fig:ldpc}(a)).
The classical capacity is given by the product of the capacities of these two: 
\begin{equation}\label{eq:C_BSEC}
    \mathcal{C}_{\text{BSEC}}=[1-h_2(p_{\mathrm{BSC}})](1-p_{\mathrm{BEC}})\;,
\end{equation}
where $h_2(x):=-x\log_2x-(1-x)\log_2(1-x)$.
Using a belief-propagation decoder~\cite{MacKay2003}, the full $\mathbf{x}_a = \mathbf{m}_a\Gb_a$ can be recovered with high probability based on the remaining bits from the corrupted $\widetilde{\mathbf{x}}_a$, which has complexity as good as $\mathcal{O}(k)$. Therefore, we can recover $\mathbf{m}_a$. Empirically, the rate of the LDPC code should be sufficiently smaller than $\mathcal{C}_{\text{BSEC}}$ of Eq.~\eqref{eq:C_BSEC}, where $p_{\mathrm{BEC}} = 1-(1-p_f)\cdot p_{X,\mathbf{b}}(\varepsilon)^2$, and $p_{X,\mathbf{b}}$ is evaluated from Eq.~\eqref{eq:pxpzR}. We simulate the logical error rate of a rate-$1/4$ regular LDPC code, whose performance is displayed in Fig.~\ref{fig:ldpc}(b,c). 
If the total bit erasure rate (the ratio of lost bits in $\mathbf{x}_a$) is about $0.55$, and the logical bit flip rate of the successful $X$ measurement is under the threshold (about $0.04$), then the information $\mathbf{m}_a$ can be recovered with an arbitrarily small error rate, if we scale up $k$. 

If one is only concerned about the bit erasure and the large $k$ limit, then the family of fountain codes is a better choice~\cite{Luby2001EECC,RaptorCodes}. These codes have well-studied decoders with complexity $\mathcal{O}(k)$ and are guaranteed to asymptotically approach the classical erasure channel capacity $C_e=1-p_{\text{BEC}}$.

\begin{thebibliography}{43}%
\makeatletter
\providecommand \@ifxundefined [1]{%
 \@ifx{#1\undefined}
}%
\providecommand \@ifnum [1]{%
 \ifnum #1\expandafter \@firstoftwo
 \else \expandafter \@secondoftwo
 \fi
}%
\providecommand \@ifx [1]{%
 \ifx #1\expandafter \@firstoftwo
 \else \expandafter \@secondoftwo
 \fi
}%
\providecommand \natexlab [1]{#1}%
\providecommand \enquote  [1]{``#1''}%
\providecommand \bibnamefont  [1]{#1}%
\providecommand \bibfnamefont [1]{#1}%
\providecommand \citenamefont [1]{#1}%
\providecommand \href@noop [0]{\@secondoftwo}%
\providecommand \href [0]{\begingroup \@sanitize@url \@href}%
\providecommand \@href[1]{\@@startlink{#1}\@@href}%
\providecommand \@@href[1]{\endgroup#1\@@endlink}%
\providecommand \@sanitize@url [0]{\catcode `\\12\catcode `\$12\catcode
  `\&12\catcode `\#12\catcode `\^12\catcode `\_12\catcode `\%12\relax}%
\providecommand \@@startlink[1]{}%
\providecommand \@@endlink[0]{}%
\providecommand \url  [0]{\begingroup\@sanitize@url \@url }%
\providecommand \@url [1]{\endgroup\@href {#1}{\urlprefix }}%
\providecommand \urlprefix  [0]{URL }%
\providecommand \Eprint [0]{\href }%
\providecommand \doibase [0]{https://doi.org/}%
\providecommand \selectlanguage [0]{\@gobble}%
\providecommand \bibinfo  [0]{\@secondoftwo}%
\providecommand \bibfield  [0]{\@secondoftwo}%
\providecommand \translation [1]{[#1]}%
\providecommand \BibitemOpen [0]{}%
\providecommand \bibitemStop [0]{}%
\providecommand \bibitemNoStop [0]{.\EOS\space}%
\providecommand \EOS [0]{\spacefactor3000\relax}%
\providecommand \BibitemShut  [1]{\csname bibitem#1\endcsname}%
\let\auto@bib@innerbib\@empty
\bibitem [{\citenamefont {Bartolucci}\ \emph {et~al.}(2023)\citenamefont
  {Bartolucci}, \citenamefont {Birchall}, \citenamefont {Bomb{\'i}n},
  \citenamefont {Cable}, \citenamefont {Dawson}, \citenamefont
  {Gimeno-Segovia}, \citenamefont {Johnston}, \citenamefont {Kieling},
  \citenamefont {Nickerson}, \citenamefont {Pant}, \citenamefont {Pastawski},
  \citenamefont {Rudolph},\ and\ \citenamefont
  {Sparrow}}]{bartolucci2021fusionbased}%
  \BibitemOpen
  \bibfield  {author} {\bibinfo {author} {\bibfnamefont {S.}~\bibnamefont
  {Bartolucci}}, \bibinfo {author} {\bibfnamefont {P.}~\bibnamefont
  {Birchall}}, \bibinfo {author} {\bibfnamefont {H.}~\bibnamefont
  {Bomb{\'i}n}}, \bibinfo {author} {\bibfnamefont {H.}~\bibnamefont {Cable}},
  \bibinfo {author} {\bibfnamefont {C.}~\bibnamefont {Dawson}}, \bibinfo
  {author} {\bibfnamefont {M.}~\bibnamefont {Gimeno-Segovia}}, \bibinfo
  {author} {\bibfnamefont {E.}~\bibnamefont {Johnston}}, \bibinfo {author}
  {\bibfnamefont {K.}~\bibnamefont {Kieling}}, \bibinfo {author} {\bibfnamefont
  {N.}~\bibnamefont {Nickerson}}, \bibinfo {author} {\bibfnamefont
  {M.}~\bibnamefont {Pant}}, \bibinfo {author} {\bibfnamefont {F.}~\bibnamefont
  {Pastawski}}, \bibinfo {author} {\bibfnamefont {T.}~\bibnamefont {Rudolph}},\
  and\ \bibinfo {author} {\bibfnamefont {C.}~\bibnamefont {Sparrow}},\
  }\bibfield  {title} {\bibinfo {title} {Fusion-based quantum computation},\
  }\href {https://doi.org/10.1038/s41467-023-36493-1} {\bibfield  {journal}
  {\bibinfo  {journal} {Nature Communications}\ }\textbf {\bibinfo {volume}
  {14}},\ \bibinfo {pages} {912} (\bibinfo {year} {2023})}\BibitemShut
  {NoStop}%
\bibitem [{\citenamefont {Kitaev}(2003)}]{Kitaev2003}%
  \BibitemOpen
  \bibfield  {author} {\bibinfo {author} {\bibfnamefont {A.}~\bibnamefont
  {Kitaev}},\ }\bibfield  {title} {\bibinfo {title} {Fault-tolerant quantum
  computation by anyons},\ }\href
  {https://doi.org/10.1016/s0003-4916(02)00018-0} {\bibfield  {journal}
  {\bibinfo  {journal} {Annals of Physics}\ }\textbf {\bibinfo {volume}
  {303}},\ \bibinfo {pages} {2} (\bibinfo {year} {2003})}\BibitemShut {NoStop}%
\bibitem [{\citenamefont {Raussendorf}\ and\ \citenamefont
  {Briegel}(2001)}]{Briegel}%
  \BibitemOpen
  \bibfield  {author} {\bibinfo {author} {\bibfnamefont {R.}~\bibnamefont
  {Raussendorf}}\ and\ \bibinfo {author} {\bibfnamefont {H.~J.}\ \bibnamefont
  {Briegel}},\ }\bibfield  {title} {\bibinfo {title} {A one-way quantum
  computer},\ }\href {https://doi.org/10.1103/PhysRevLett.86.5188} {\bibfield
  {journal} {\bibinfo  {journal} {Phys. Rev. Lett.}\ }\textbf {\bibinfo
  {volume} {86}},\ \bibinfo {pages} {5188} (\bibinfo {year}
  {2001})}\BibitemShut {NoStop}%
\bibitem [{\citenamefont {Sangouard}\ \emph {et~al.}(2011)\citenamefont
  {Sangouard}, \citenamefont {Simon}, \citenamefont {de~Riedmatten},\ and\
  \citenamefont {Gisin}}]{Sangouard2011}%
  \BibitemOpen
  \bibfield  {author} {\bibinfo {author} {\bibfnamefont {N.}~\bibnamefont
  {Sangouard}}, \bibinfo {author} {\bibfnamefont {C.}~\bibnamefont {Simon}},
  \bibinfo {author} {\bibfnamefont {H.}~\bibnamefont {de~Riedmatten}},\ and\
  \bibinfo {author} {\bibfnamefont {N.}~\bibnamefont {Gisin}},\ }\bibfield
  {title} {\bibinfo {title} {Quantum repeaters based on atomic ensembles and
  linear optics},\ }\href {https://doi.org/10.1103/RevModPhys.83.33} {\bibfield
   {journal} {\bibinfo  {journal} {Rev. Mod. Phys.}\ }\textbf {\bibinfo
  {volume} {83}},\ \bibinfo {pages} {33} (\bibinfo {year} {2011})}\BibitemShut
  {NoStop}%
\bibitem [{\citenamefont {Azuma}\ \emph {et~al.}(2015)\citenamefont {Azuma},
  \citenamefont {Tamaki},\ and\ \citenamefont {Lo}}]{Azuma2015}%
  \BibitemOpen
  \bibfield  {author} {\bibinfo {author} {\bibfnamefont {K.}~\bibnamefont
  {Azuma}}, \bibinfo {author} {\bibfnamefont {K.}~\bibnamefont {Tamaki}},\ and\
  \bibinfo {author} {\bibfnamefont {H.-K.}\ \bibnamefont {Lo}},\ }\bibfield
  {title} {\bibinfo {title} {All-photonic quantum repeaters},\ }\href
  {https://doi.org/10.1038/ncomms7787} {\bibfield  {journal} {\bibinfo
  {journal} {Nature Communications}\ }\textbf {\bibinfo {volume} {6}},\
  \bibinfo {pages} {6787} (\bibinfo {year} {2015})}\BibitemShut {NoStop}%
\bibitem [{\citenamefont {Shettell}\ and\ \citenamefont
  {Markham}(2020)}]{Damian2020}%
  \BibitemOpen
  \bibfield  {author} {\bibinfo {author} {\bibfnamefont {N.}~\bibnamefont
  {Shettell}}\ and\ \bibinfo {author} {\bibfnamefont {D.}~\bibnamefont
  {Markham}},\ }\bibfield  {title} {\bibinfo {title} {Graph states as a
  resource for quantum metrology},\ }\href
  {https://doi.org/10.1103/PhysRevLett.124.110502} {\bibfield  {journal}
  {\bibinfo  {journal} {Phys. Rev. Lett.}\ }\textbf {\bibinfo {volume} {124}},\
  \bibinfo {pages} {110502} (\bibinfo {year} {2020})}\BibitemShut {NoStop}%
\bibitem [{\citenamefont {Bennett}\ \emph {et~al.}(1992)\citenamefont
  {Bennett}, \citenamefont {Brassard},\ and\ \citenamefont {Mermin}}]{E92}%
  \BibitemOpen
  \bibfield  {author} {\bibinfo {author} {\bibfnamefont {C.~H.}\ \bibnamefont
  {Bennett}}, \bibinfo {author} {\bibfnamefont {G.}~\bibnamefont {Brassard}},\
  and\ \bibinfo {author} {\bibfnamefont {N.~D.}\ \bibnamefont {Mermin}},\
  }\bibfield  {title} {\bibinfo {title} {Quantum cryptography without bell's
  theorem},\ }\href {https://doi.org/10.1103/PhysRevLett.68.557} {\bibfield
  {journal} {\bibinfo  {journal} {Phys. Rev. Lett.}\ }\textbf {\bibinfo
  {volume} {68}},\ \bibinfo {pages} {557} (\bibinfo {year} {1992})}\BibitemShut
  {NoStop}%
\bibitem [{\citenamefont {Muralidharan}\ \emph {et~al.}(2016)\citenamefont
  {Muralidharan}, \citenamefont {Li}, \citenamefont {Kim}, \citenamefont
  {L{\"u}tkenhaus}, \citenamefont {Lukin},\ and\ \citenamefont
  {Jiang}}]{Muralidharan2016}%
  \BibitemOpen
  \bibfield  {author} {\bibinfo {author} {\bibfnamefont {S.}~\bibnamefont
  {Muralidharan}}, \bibinfo {author} {\bibfnamefont {L.}~\bibnamefont {Li}},
  \bibinfo {author} {\bibfnamefont {J.}~\bibnamefont {Kim}}, \bibinfo {author}
  {\bibfnamefont {N.}~\bibnamefont {L{\"u}tkenhaus}}, \bibinfo {author}
  {\bibfnamefont {M.~D.}\ \bibnamefont {Lukin}},\ and\ \bibinfo {author}
  {\bibfnamefont {L.}~\bibnamefont {Jiang}},\ }\bibfield  {title} {\bibinfo
  {title} {Optimal architectures for long distance quantum communication},\
  }\href {https://doi.org/10.1038/srep20463} {\bibfield  {journal} {\bibinfo
  {journal} {Scientific Reports}\ }\textbf {\bibinfo {volume} {6}},\ \bibinfo
  {pages} {20463} (\bibinfo {year} {2016})}\BibitemShut {NoStop}%
\bibitem [{\citenamefont {Azuma}\ \emph {et~al.}(2023)\citenamefont {Azuma},
  \citenamefont {Economou}, \citenamefont {Elkouss}, \citenamefont {Hilaire},
  \citenamefont {Jiang}, \citenamefont {Lo},\ and\ \citenamefont
  {Tzitrin}}]{RevModPhys.95.045006}%
  \BibitemOpen
  \bibfield  {author} {\bibinfo {author} {\bibfnamefont {K.}~\bibnamefont
  {Azuma}}, \bibinfo {author} {\bibfnamefont {S.~E.}\ \bibnamefont {Economou}},
  \bibinfo {author} {\bibfnamefont {D.}~\bibnamefont {Elkouss}}, \bibinfo
  {author} {\bibfnamefont {P.}~\bibnamefont {Hilaire}}, \bibinfo {author}
  {\bibfnamefont {L.}~\bibnamefont {Jiang}}, \bibinfo {author} {\bibfnamefont
  {H.-K.}\ \bibnamefont {Lo}},\ and\ \bibinfo {author} {\bibfnamefont
  {I.}~\bibnamefont {Tzitrin}},\ }\bibfield  {title} {\bibinfo {title} {Quantum
  repeaters: From quantum networks to the quantum internet},\ }\href
  {https://doi.org/10.1103/RevModPhys.95.045006} {\bibfield  {journal}
  {\bibinfo  {journal} {Rev. Mod. Phys.}\ }\textbf {\bibinfo {volume} {95}},\
  \bibinfo {pages} {045006} (\bibinfo {year} {2023})}\BibitemShut {NoStop}%
\bibitem [{\citenamefont {Munro}\ \emph {et~al.}(2015)\citenamefont {Munro},
  \citenamefont {Azuma}, \citenamefont {Tamaki},\ and\ \citenamefont
  {Nemoto}}]{Munro2015review}%
  \BibitemOpen
  \bibfield  {author} {\bibinfo {author} {\bibfnamefont {W.~J.}\ \bibnamefont
  {Munro}}, \bibinfo {author} {\bibfnamefont {K.}~\bibnamefont {Azuma}},
  \bibinfo {author} {\bibfnamefont {K.}~\bibnamefont {Tamaki}},\ and\ \bibinfo
  {author} {\bibfnamefont {K.}~\bibnamefont {Nemoto}},\ }\bibfield  {title}
  {\bibinfo {title} {Inside quantum repeaters},\ }\href
  {https://doi.org/10.1109/JSTQE.2015.2392076} {\bibfield  {journal} {\bibinfo
  {journal} {IEEE Journal of Selected Topics in Quantum Electronics}\ }\textbf
  {\bibinfo {volume} {21}},\ \bibinfo {pages} {78} (\bibinfo {year}
  {2015})}\BibitemShut {NoStop}%
\bibitem [{\citenamefont {Rozp\k{e}dek}\ \emph {et~al.}(2023)\citenamefont
  {Rozp\k{e}dek}, \citenamefont {Seshadreesan}, \citenamefont {Polakos},
  \citenamefont {Jiang},\ and\ \citenamefont
  {Guha}}]{PhysRevResearch.5.043056}%
  \BibitemOpen
  \bibfield  {author} {\bibinfo {author} {\bibfnamefont {F.}~\bibnamefont
  {Rozp\k{e}dek}}, \bibinfo {author} {\bibfnamefont {K.~P.}\ \bibnamefont
  {Seshadreesan}}, \bibinfo {author} {\bibfnamefont {P.}~\bibnamefont
  {Polakos}}, \bibinfo {author} {\bibfnamefont {L.}~\bibnamefont {Jiang}},\
  and\ \bibinfo {author} {\bibfnamefont {S.}~\bibnamefont {Guha}},\ }\bibfield
  {title} {\bibinfo {title} {All-photonic gottesman-kitaev-preskill--qubit
  repeater using analog-information-assisted multiplexed entanglement
  ranking},\ }\href {https://doi.org/10.1103/PhysRevResearch.5.043056}
  {\bibfield  {journal} {\bibinfo  {journal} {Phys. Rev. Res.}\ }\textbf
  {\bibinfo {volume} {5}},\ \bibinfo {pages} {043056} (\bibinfo {year}
  {2023})}\BibitemShut {NoStop}%
\bibitem [{\citenamefont {Pant}\ \emph {et~al.}(2017)\citenamefont {Pant},
  \citenamefont {Krovi}, \citenamefont {Englund},\ and\ \citenamefont
  {Guha}}]{PhysRevA.95.012304}%
  \BibitemOpen
  \bibfield  {author} {\bibinfo {author} {\bibfnamefont {M.}~\bibnamefont
  {Pant}}, \bibinfo {author} {\bibfnamefont {H.}~\bibnamefont {Krovi}},
  \bibinfo {author} {\bibfnamefont {D.}~\bibnamefont {Englund}},\ and\ \bibinfo
  {author} {\bibfnamefont {S.}~\bibnamefont {Guha}},\ }\bibfield  {title}
  {\bibinfo {title} {Rate-distance tradeoff and resource costs for all-optical
  quantum repeaters},\ }\href {https://doi.org/10.1103/PhysRevA.95.012304}
  {\bibfield  {journal} {\bibinfo  {journal} {Phys. Rev. A}\ }\textbf {\bibinfo
  {volume} {95}},\ \bibinfo {pages} {012304} (\bibinfo {year}
  {2017})}\BibitemShut {NoStop}%
\bibitem [{\citenamefont {Ewert}\ and\ \citenamefont {van
  Loock}(2014)}]{Ewert2014PRL}%
  \BibitemOpen
  \bibfield  {author} {\bibinfo {author} {\bibfnamefont {F.}~\bibnamefont
  {Ewert}}\ and\ \bibinfo {author} {\bibfnamefont {P.}~\bibnamefont {van
  Loock}},\ }\bibfield  {title} {\bibinfo {title} {$3/4$-efficient bell
  measurement with passive linear optics and unentangled ancillae},\ }\href
  {https://doi.org/10.1103/PhysRevLett.113.140403} {\bibfield  {journal}
  {\bibinfo  {journal} {Phys. Rev. Lett.}\ }\textbf {\bibinfo {volume} {113}},\
  \bibinfo {pages} {140403} (\bibinfo {year} {2014})}\BibitemShut {NoStop}%
\bibitem [{\citenamefont {Ewert}\ \emph {et~al.}(2016)\citenamefont {Ewert},
  \citenamefont {Bergmann},\ and\ \citenamefont {van Loock}}]{Ewert2016PRL}%
  \BibitemOpen
  \bibfield  {author} {\bibinfo {author} {\bibfnamefont {F.}~\bibnamefont
  {Ewert}}, \bibinfo {author} {\bibfnamefont {M.}~\bibnamefont {Bergmann}},\
  and\ \bibinfo {author} {\bibfnamefont {P.}~\bibnamefont {van Loock}},\
  }\bibfield  {title} {\bibinfo {title} {Ultrafast long-distance quantum
  communication with static linear optics},\ }\href
  {https://doi.org/10.1103/PhysRevLett.117.210501} {\bibfield  {journal}
  {\bibinfo  {journal} {Phys. Rev. Lett.}\ }\textbf {\bibinfo {volume} {117}},\
  \bibinfo {pages} {210501} (\bibinfo {year} {2016})}\BibitemShut {NoStop}%
\bibitem [{\citenamefont {Fukui}\ \emph {et~al.}(2021)\citenamefont {Fukui},
  \citenamefont {Alexander},\ and\ \citenamefont {van Loock}}]{fukui2021all}%
  \BibitemOpen
  \bibfield  {author} {\bibinfo {author} {\bibfnamefont {K.}~\bibnamefont
  {Fukui}}, \bibinfo {author} {\bibfnamefont {R.~N.}\ \bibnamefont
  {Alexander}},\ and\ \bibinfo {author} {\bibfnamefont {P.}~\bibnamefont {van
  Loock}},\ }\bibfield  {title} {\bibinfo {title} {All-optical long-distance
  quantum communication with gottesman-kitaev-preskill qubits},\ }\href@noop {}
  {\bibfield  {journal} {\bibinfo  {journal} {Physical Review Research}\
  }\textbf {\bibinfo {volume} {3}},\ \bibinfo {pages} {033118} (\bibinfo {year}
  {2021})}\BibitemShut {NoStop}%
\bibitem [{\citenamefont {Niu}\ \emph {et~al.}(2022)\citenamefont {Niu},
  \citenamefont {Zhang}, \citenamefont {Shabani},\ and\ \citenamefont
  {Shapourian}}]{niu2022all}%
  \BibitemOpen
  \bibfield  {author} {\bibinfo {author} {\bibfnamefont {D.}~\bibnamefont
  {Niu}}, \bibinfo {author} {\bibfnamefont {Y.}~\bibnamefont {Zhang}}, \bibinfo
  {author} {\bibfnamefont {A.}~\bibnamefont {Shabani}},\ and\ \bibinfo {author}
  {\bibfnamefont {H.}~\bibnamefont {Shapourian}},\ }\bibfield  {title}
  {\bibinfo {title} {All-photonic one-way quantum repeaters},\ }\href@noop {}
  {\bibfield  {journal} {\bibinfo  {journal} {arXiv preprint arXiv:2210.10071}\
  } (\bibinfo {year} {2022})}\BibitemShut {NoStop}%
\bibitem [{\citenamefont {Rozp{\k{e}}dek}\ \emph {et~al.}(2023)\citenamefont
  {Rozp{\k{e}}dek}, \citenamefont {Seshadreesan}, \citenamefont {Polakos},
  \citenamefont {Jiang},\ and\ \citenamefont {Guha}}]{rozpkedek2023all}%
  \BibitemOpen
  \bibfield  {author} {\bibinfo {author} {\bibfnamefont {F.}~\bibnamefont
  {Rozp{\k{e}}dek}}, \bibinfo {author} {\bibfnamefont {K.~P.}\ \bibnamefont
  {Seshadreesan}}, \bibinfo {author} {\bibfnamefont {P.}~\bibnamefont
  {Polakos}}, \bibinfo {author} {\bibfnamefont {L.}~\bibnamefont {Jiang}},\
  and\ \bibinfo {author} {\bibfnamefont {S.}~\bibnamefont {Guha}},\ }\bibfield
  {title} {\bibinfo {title} {All-photonic gottesman-kitaev-preskill--qubit
  repeater using analog-information-assisted multiplexed entanglement
  ranking},\ }\href@noop {} {\bibfield  {journal} {\bibinfo  {journal}
  {Physical Review Research}\ }\textbf {\bibinfo {volume} {5}},\ \bibinfo
  {pages} {043056} (\bibinfo {year} {2023})}\BibitemShut {NoStop}%
\bibitem [{\citenamefont {Patil}\ and\ \citenamefont
  {Guha}(2024)}]{patil2024improved}%
  \BibitemOpen
  \bibfield  {author} {\bibinfo {author} {\bibfnamefont {A.}~\bibnamefont
  {Patil}}\ and\ \bibinfo {author} {\bibfnamefont {S.}~\bibnamefont {Guha}},\
  }\href@noop {} {\bibinfo {title} {An improved design for all-photonic quantum
  repeaters}} (\bibinfo {year} {2024}),\ \Eprint
  {https://arxiv.org/abs/2405.11768} {arXiv:2405.11768 [quant-ph]} \BibitemShut
  {NoStop}%
\bibitem [{\citenamefont {Browne}\ and\ \citenamefont
  {Rudolph}(2005)}]{fusiongates}%
  \BibitemOpen
  \bibfield  {author} {\bibinfo {author} {\bibfnamefont {D.~E.}\ \bibnamefont
  {Browne}}\ and\ \bibinfo {author} {\bibfnamefont {T.}~\bibnamefont
  {Rudolph}},\ }\bibfield  {title} {\bibinfo {title} {Resource-efficient linear
  optical quantum computation},\ }\href
  {https://doi.org/10.1103/PhysRevLett.95.010501} {\bibfield  {journal}
  {\bibinfo  {journal} {Phys. Rev. Lett.}\ }\textbf {\bibinfo {volume} {95}},\
  \bibinfo {pages} {010501} (\bibinfo {year} {2005})}\BibitemShut {NoStop}%
\bibitem [{\citenamefont {L\"utkenhaus}\ \emph {et~al.}(1999)\citenamefont
  {L\"utkenhaus}, \citenamefont {Calsamiglia},\ and\ \citenamefont
  {Suominen}}]{nogo_deterministic_BSM}%
  \BibitemOpen
  \bibfield  {author} {\bibinfo {author} {\bibfnamefont {N.}~\bibnamefont
  {L\"utkenhaus}}, \bibinfo {author} {\bibfnamefont {J.}~\bibnamefont
  {Calsamiglia}},\ and\ \bibinfo {author} {\bibfnamefont {K.-A.}\ \bibnamefont
  {Suominen}},\ }\bibfield  {title} {\bibinfo {title} {Bell measurements for
  teleportation},\ }\href {https://doi.org/10.1103/PhysRevA.59.3295} {\bibfield
   {journal} {\bibinfo  {journal} {Phys. Rev. A}\ }\textbf {\bibinfo {volume}
  {59}},\ \bibinfo {pages} {3295} (\bibinfo {year} {1999})}\BibitemShut
  {NoStop}%
\bibitem [{\citenamefont {Zwerger}\ \emph {et~al.}(2012)\citenamefont
  {Zwerger}, \citenamefont {D\"ur},\ and\ \citenamefont
  {Briegel}}]{Zwerger2012PRA}%
  \BibitemOpen
  \bibfield  {author} {\bibinfo {author} {\bibfnamefont {M.}~\bibnamefont
  {Zwerger}}, \bibinfo {author} {\bibfnamefont {W.}~\bibnamefont {D\"ur}},\
  and\ \bibinfo {author} {\bibfnamefont {H.~J.}\ \bibnamefont {Briegel}},\
  }\bibfield  {title} {\bibinfo {title} {Measurement-based quantum repeaters},\
  }\href {https://doi.org/10.1103/PhysRevA.85.062326} {\bibfield  {journal}
  {\bibinfo  {journal} {Phys. Rev. A}\ }\textbf {\bibinfo {volume} {85}},\
  \bibinfo {pages} {062326} (\bibinfo {year} {2012})}\BibitemShut {NoStop}%
\bibitem [{\citenamefont {Zwerger}\ \emph {et~al.}(2018)\citenamefont
  {Zwerger}, \citenamefont {Pirker}, \citenamefont {Dunjko}, \citenamefont
  {Briegel},\ and\ \citenamefont {D\"ur}}]{Zwerger2018PRL}%
  \BibitemOpen
  \bibfield  {author} {\bibinfo {author} {\bibfnamefont {M.}~\bibnamefont
  {Zwerger}}, \bibinfo {author} {\bibfnamefont {A.}~\bibnamefont {Pirker}},
  \bibinfo {author} {\bibfnamefont {V.}~\bibnamefont {Dunjko}}, \bibinfo
  {author} {\bibfnamefont {H.~J.}\ \bibnamefont {Briegel}},\ and\ \bibinfo
  {author} {\bibfnamefont {W.}~\bibnamefont {D\"ur}},\ }\bibfield  {title}
  {\bibinfo {title} {Long-range big quantum-data transmission},\ }\href
  {https://doi.org/10.1103/PhysRevLett.120.030503} {\bibfield  {journal}
  {\bibinfo  {journal} {Phys. Rev. Lett.}\ }\textbf {\bibinfo {volume} {120}},\
  \bibinfo {pages} {030503} (\bibinfo {year} {2018})}\BibitemShut {NoStop}%
\bibitem [{\citenamefont {Bennett}\ \emph {et~al.}(1997)\citenamefont
  {Bennett}, \citenamefont {DiVincenzo},\ and\ \citenamefont
  {Smolin}}]{ErasChnCapacity1997PRL}%
  \BibitemOpen
  \bibfield  {author} {\bibinfo {author} {\bibfnamefont {C.~H.}\ \bibnamefont
  {Bennett}}, \bibinfo {author} {\bibfnamefont {D.~P.}\ \bibnamefont
  {DiVincenzo}},\ and\ \bibinfo {author} {\bibfnamefont {J.~A.}\ \bibnamefont
  {Smolin}},\ }\bibfield  {title} {\bibinfo {title} {Capacities of quantum
  erasure channels},\ }\href {https://doi.org/10.1103/PhysRevLett.78.3217}
  {\bibfield  {journal} {\bibinfo  {journal} {Phys. Rev. Lett.}\ }\textbf
  {\bibinfo {volume} {78}},\ \bibinfo {pages} {3217} (\bibinfo {year}
  {1997})}\BibitemShut {NoStop}%
\bibitem [{\citenamefont {Li}\ \emph {et~al.}(2019)\citenamefont {Li},
  \citenamefont {Zhang}, \citenamefont {Yin}, \citenamefont {Liu},
  \citenamefont {Hu}, \citenamefont {Fang}, \citenamefont {Fei}, \citenamefont
  {Jiang}, \citenamefont {Zhang}, \citenamefont {Li}, \citenamefont {Liu},
  \citenamefont {Xu}, \citenamefont {Chen},\ and\ \citenamefont
  {Pan}}]{Li2019}%
  \BibitemOpen
  \bibfield  {author} {\bibinfo {author} {\bibfnamefont {Z.-D.}\ \bibnamefont
  {Li}}, \bibinfo {author} {\bibfnamefont {R.}~\bibnamefont {Zhang}}, \bibinfo
  {author} {\bibfnamefont {X.-F.}\ \bibnamefont {Yin}}, \bibinfo {author}
  {\bibfnamefont {L.-Z.}\ \bibnamefont {Liu}}, \bibinfo {author} {\bibfnamefont
  {Y.}~\bibnamefont {Hu}}, \bibinfo {author} {\bibfnamefont {Y.-Q.}\
  \bibnamefont {Fang}}, \bibinfo {author} {\bibfnamefont {Y.-Y.}\ \bibnamefont
  {Fei}}, \bibinfo {author} {\bibfnamefont {X.}~\bibnamefont {Jiang}}, \bibinfo
  {author} {\bibfnamefont {J.}~\bibnamefont {Zhang}}, \bibinfo {author}
  {\bibfnamefont {L.}~\bibnamefont {Li}}, \bibinfo {author} {\bibfnamefont
  {N.-L.}\ \bibnamefont {Liu}}, \bibinfo {author} {\bibfnamefont
  {F.}~\bibnamefont {Xu}}, \bibinfo {author} {\bibfnamefont {Y.-A.}\
  \bibnamefont {Chen}},\ and\ \bibinfo {author} {\bibfnamefont {J.-W.}\
  \bibnamefont {Pan}},\ }\bibfield  {title} {\bibinfo {title} {Experimental
  quantum repeater without quantum memory},\ }\href
  {https://doi.org/10.1038/s41566-019-0468-5} {\bibfield  {journal} {\bibinfo
  {journal} {Nature Photonics}\ }\textbf {\bibinfo {volume} {13}},\ \bibinfo
  {pages} {644} (\bibinfo {year} {2019})}\BibitemShut {NoStop}%
\bibitem [{\citenamefont {Buterakos}\ \emph {et~al.}(2017)\citenamefont
  {Buterakos}, \citenamefont {Barnes},\ and\ \citenamefont
  {Economou}}]{ButerakosPRX2017}%
  \BibitemOpen
  \bibfield  {author} {\bibinfo {author} {\bibfnamefont {D.}~\bibnamefont
  {Buterakos}}, \bibinfo {author} {\bibfnamefont {E.}~\bibnamefont {Barnes}},\
  and\ \bibinfo {author} {\bibfnamefont {S.~E.}\ \bibnamefont {Economou}},\
  }\bibfield  {title} {\bibinfo {title} {Deterministic generation of
  all-photonic quantum repeaters from solid-state emitters},\ }\href
  {https://doi.org/10.1103/PhysRevX.7.041023} {\bibfield  {journal} {\bibinfo
  {journal} {Phys. Rev. X}\ }\textbf {\bibinfo {volume} {7}},\ \bibinfo {pages}
  {041023} (\bibinfo {year} {2017})}\BibitemShut {NoStop}%
\bibitem [{\citenamefont {Zhan}\ \emph {et~al.}(2023)\citenamefont {Zhan},
  \citenamefont {Hilaire}, \citenamefont {Barnes}, \citenamefont {Economou},\
  and\ \citenamefont {Sun}}]{Zhan2023performanceanalysis}%
  \BibitemOpen
  \bibfield  {author} {\bibinfo {author} {\bibfnamefont {Y.}~\bibnamefont
  {Zhan}}, \bibinfo {author} {\bibfnamefont {P.}~\bibnamefont {Hilaire}},
  \bibinfo {author} {\bibfnamefont {E.}~\bibnamefont {Barnes}}, \bibinfo
  {author} {\bibfnamefont {S.~E.}\ \bibnamefont {Economou}},\ and\ \bibinfo
  {author} {\bibfnamefont {S.}~\bibnamefont {Sun}},\ }\bibfield  {title}
  {\bibinfo {title} {Performance analysis of quantum repeaters enabled by
  deterministically generated photonic graph states},\ }\href
  {https://doi.org/10.22331/q-2023-02-16-924} {\bibfield  {journal} {\bibinfo
  {journal} {{Quantum}}\ }\textbf {\bibinfo {volume} {7}},\ \bibinfo {pages}
  {924} (\bibinfo {year} {2023})}\BibitemShut {NoStop}%
\bibitem [{\citenamefont {MacKay}(2003)}]{MacKay2003}%
  \BibitemOpen
  \bibfield  {author} {\bibinfo {author} {\bibfnamefont {D.~J.~C.}\
  \bibnamefont {MacKay}},\ }\href@noop {} {\emph {\bibinfo {title} {Information
  Theory, Inference, and Learning Algorithms}}}\ (\bibinfo  {publisher}
  {Copyright Cambridge University Press},\ \bibinfo {year} {2003})\BibitemShut
  {NoStop}%
\bibitem [{\citenamefont {Luby}\ \emph {et~al.}(2001)\citenamefont {Luby},
  \citenamefont {Mitzenmacher}, \citenamefont {Shokrollahi},\ and\
  \citenamefont {Spielman}}]{Luby2001EECC}%
  \BibitemOpen
  \bibfield  {author} {\bibinfo {author} {\bibfnamefont {M.}~\bibnamefont
  {Luby}}, \bibinfo {author} {\bibfnamefont {M.}~\bibnamefont {Mitzenmacher}},
  \bibinfo {author} {\bibfnamefont {M.}~\bibnamefont {Shokrollahi}},\ and\
  \bibinfo {author} {\bibfnamefont {D.}~\bibnamefont {Spielman}},\ }\bibfield
  {title} {\bibinfo {title} {Efficient erasure correcting codes},\ }\href
  {https://doi.org/10.1109/18.910575} {\bibfield  {journal} {\bibinfo
  {journal} {IEEE Transactions on Information Theory}\ }\textbf {\bibinfo
  {volume} {47}},\ \bibinfo {pages} {569} (\bibinfo {year} {2001})}\BibitemShut
  {NoStop}%
\bibitem [{\citenamefont {Shokrollahi}(2006)}]{RaptorCodes}%
  \BibitemOpen
  \bibfield  {author} {\bibinfo {author} {\bibfnamefont {A.}~\bibnamefont
  {Shokrollahi}},\ }\bibfield  {title} {\bibinfo {title} {Raptor codes},\
  }\href {https://doi.org/10.1109/TIT.2006.874390} {\bibfield  {journal}
  {\bibinfo  {journal} {IEEE Transactions on Information Theory}\ }\textbf
  {\bibinfo {volume} {52}},\ \bibinfo {pages} {2551} (\bibinfo {year}
  {2006})}\BibitemShut {NoStop}%
\bibitem [{\citenamefont {Varnava}\ \emph {et~al.}(2006)\citenamefont
  {Varnava}, \citenamefont {Browne},\ and\ \citenamefont
  {Rudolph}}]{Varnava2006}%
  \BibitemOpen
  \bibfield  {author} {\bibinfo {author} {\bibfnamefont {M.}~\bibnamefont
  {Varnava}}, \bibinfo {author} {\bibfnamefont {D.~E.}\ \bibnamefont
  {Browne}},\ and\ \bibinfo {author} {\bibfnamefont {T.}~\bibnamefont
  {Rudolph}},\ }\bibfield  {title} {\bibinfo {title} {Loss tolerance in one-way
  quantum computation via counterfactual error correction},\ }\href
  {https://doi.org/10.1103/PhysRevLett.97.120501} {\bibfield  {journal}
  {\bibinfo  {journal} {Phys. Rev. Lett.}\ }\textbf {\bibinfo {volume} {97}},\
  \bibinfo {pages} {120501} (\bibinfo {year} {2006})}\BibitemShut {NoStop}%
\bibitem [{Sup()}]{SuppMat}%
  \BibitemOpen
  \href@noop {} {\bibinfo {title} {See supplemental material for technical
  details and additional examples.}}\BibitemShut {Stop}%
\bibitem [{\citenamefont {Hein}\ \emph {et~al.}(2004)\citenamefont {Hein},
  \citenamefont {Eisert},\ and\ \citenamefont {Briegel}}]{Hein2004PRA}%
  \BibitemOpen
  \bibfield  {author} {\bibinfo {author} {\bibfnamefont {M.}~\bibnamefont
  {Hein}}, \bibinfo {author} {\bibfnamefont {J.}~\bibnamefont {Eisert}},\ and\
  \bibinfo {author} {\bibfnamefont {H.~J.}\ \bibnamefont {Briegel}},\
  }\bibfield  {title} {\bibinfo {title} {Multiparty entanglement in graph
  states},\ }\href {https://doi.org/10.1103/PhysRevA.69.062311} {\bibfield
  {journal} {\bibinfo  {journal} {Phys. Rev. A}\ }\textbf {\bibinfo {volume}
  {69}},\ \bibinfo {pages} {062311} (\bibinfo {year} {2004})}\BibitemShut
  {NoStop}%
\bibitem [{\citenamefont {Tzitrin}(2018)}]{Tzitrin2018PRA}%
  \BibitemOpen
  \bibfield  {author} {\bibinfo {author} {\bibfnamefont {I.}~\bibnamefont
  {Tzitrin}},\ }\bibfield  {title} {\bibinfo {title} {Local equivalence of
  complete bipartite and repeater graph states},\ }\href
  {https://doi.org/10.1103/PhysRevA.98.032305} {\bibfield  {journal} {\bibinfo
  {journal} {Phys. Rev. A}\ }\textbf {\bibinfo {volume} {98}},\ \bibinfo
  {pages} {032305} (\bibinfo {year} {2018})}\BibitemShut {NoStop}%
\bibitem [{\citenamefont {Russo}\ \emph {et~al.}(2018)\citenamefont {Russo},
  \citenamefont {Barnes},\ and\ \citenamefont {Economou}}]{Russo2018PRB}%
  \BibitemOpen
  \bibfield  {author} {\bibinfo {author} {\bibfnamefont {A.}~\bibnamefont
  {Russo}}, \bibinfo {author} {\bibfnamefont {E.}~\bibnamefont {Barnes}},\ and\
  \bibinfo {author} {\bibfnamefont {S.~E.}\ \bibnamefont {Economou}},\
  }\bibfield  {title} {\bibinfo {title} {Photonic graph state generation from
  quantum dots and color centers for quantum communications},\ }\href
  {https://doi.org/10.1103/PhysRevB.98.085303} {\bibfield  {journal} {\bibinfo
  {journal} {Phys. Rev. B}\ }\textbf {\bibinfo {volume} {98}},\ \bibinfo
  {pages} {085303} (\bibinfo {year} {2018})}\BibitemShut {NoStop}%
\bibitem [{\citenamefont {Gallager}(1962)}]{Gallager_LDPC}%
  \BibitemOpen
  \bibfield  {author} {\bibinfo {author} {\bibfnamefont {R.}~\bibnamefont
  {Gallager}},\ }\bibfield  {title} {\bibinfo {title} {Low-density parity-check
  codes},\ }\href {https://doi.org/10.1109/TIT.1962.1057683} {\bibfield
  {journal} {\bibinfo  {journal} {IRE Transactions on Information Theory}\
  }\textbf {\bibinfo {volume} {8}},\ \bibinfo {pages} {21} (\bibinfo {year}
  {1962})}\BibitemShut {NoStop}%
\bibitem [{\citenamefont {Economou}\ \emph {et~al.}(2010)\citenamefont
  {Economou}, \citenamefont {Lindner},\ and\ \citenamefont
  {Rudolph}}]{Economou2010}%
  \BibitemOpen
  \bibfield  {author} {\bibinfo {author} {\bibfnamefont {S.~E.}\ \bibnamefont
  {Economou}}, \bibinfo {author} {\bibfnamefont {N.}~\bibnamefont {Lindner}},\
  and\ \bibinfo {author} {\bibfnamefont {T.}~\bibnamefont {Rudolph}},\
  }\bibfield  {title} {\bibinfo {title} {Optically generated 2-dimensional
  photonic cluster state from coupled quantum dots},\ }\href
  {https://doi.org/10.1103/PhysRevLett.105.093601} {\bibfield  {journal}
  {\bibinfo  {journal} {Phys. Rev. Lett.}\ }\textbf {\bibinfo {volume} {105}},\
  \bibinfo {pages} {093601} (\bibinfo {year} {2010})}\BibitemShut {NoStop}%
\bibitem [{\citenamefont {Pichler}\ \emph {et~al.}(2017)\citenamefont
  {Pichler}, \citenamefont {Choi}, \citenamefont {Zoller},\ and\ \citenamefont
  {Lukin}}]{Pichler2017}%
  \BibitemOpen
  \bibfield  {author} {\bibinfo {author} {\bibfnamefont {H.}~\bibnamefont
  {Pichler}}, \bibinfo {author} {\bibfnamefont {S.}~\bibnamefont {Choi}},
  \bibinfo {author} {\bibfnamefont {P.}~\bibnamefont {Zoller}},\ and\ \bibinfo
  {author} {\bibfnamefont {M.~D.}\ \bibnamefont {Lukin}},\ }\bibfield  {title}
  {\bibinfo {title} {Universal photonic quantum computation via time-delayed
  feedback},\ }\href {https://doi.org/10.1073/pnas.1711003114} {\bibfield
  {journal} {\bibinfo  {journal} {Proceedings of the National Academy of
  Sciences}\ }\textbf {\bibinfo {volume} {114}},\ \bibinfo {pages} {11362}
  (\bibinfo {year} {2017})}\BibitemShut {NoStop}%
\bibitem [{\citenamefont {Schwartz}\ \emph {et~al.}(2016)\citenamefont
  {Schwartz}, \citenamefont {Cogan}, \citenamefont {Schmidgall}, \citenamefont
  {Don}, \citenamefont {Gantz}, \citenamefont {Kenneth}, \citenamefont
  {Lindner},\ and\ \citenamefont {Gershoni}}]{Schwartz2016}%
  \BibitemOpen
  \bibfield  {author} {\bibinfo {author} {\bibfnamefont {I.}~\bibnamefont
  {Schwartz}}, \bibinfo {author} {\bibfnamefont {D.}~\bibnamefont {Cogan}},
  \bibinfo {author} {\bibfnamefont {E.~R.}\ \bibnamefont {Schmidgall}},
  \bibinfo {author} {\bibfnamefont {Y.}~\bibnamefont {Don}}, \bibinfo {author}
  {\bibfnamefont {L.}~\bibnamefont {Gantz}}, \bibinfo {author} {\bibfnamefont
  {O.}~\bibnamefont {Kenneth}}, \bibinfo {author} {\bibfnamefont {N.~H.}\
  \bibnamefont {Lindner}},\ and\ \bibinfo {author} {\bibfnamefont
  {D.}~\bibnamefont {Gershoni}},\ }\bibfield  {title} {\bibinfo {title}
  {Deterministic generation of a cluster state of entangled photons},\ }\href
  {https://doi.org/10.1126/science.aah4758} {\bibfield  {journal} {\bibinfo
  {journal} {Science}\ }\textbf {\bibinfo {volume} {354}},\ \bibinfo {pages}
  {434} (\bibinfo {year} {2016})}\BibitemShut {NoStop}%
\bibitem [{\citenamefont {Besse}\ \emph {et~al.}(2020)\citenamefont {Besse},
  \citenamefont {Reuer}, \citenamefont {Collodo}, \citenamefont {Wulff},
  \citenamefont {Wernli}, \citenamefont {Copetudo}, \citenamefont {Malz},
  \citenamefont {Magnard}, \citenamefont {Akin}, \citenamefont {Gabureac},
  \citenamefont {Norris}, \citenamefont {Cirac}, \citenamefont {Wallraff},\
  and\ \citenamefont {Eichler}}]{Besse2020}%
  \BibitemOpen
  \bibfield  {author} {\bibinfo {author} {\bibfnamefont {J.-C.}\ \bibnamefont
  {Besse}}, \bibinfo {author} {\bibfnamefont {K.}~\bibnamefont {Reuer}},
  \bibinfo {author} {\bibfnamefont {M.~C.}\ \bibnamefont {Collodo}}, \bibinfo
  {author} {\bibfnamefont {A.}~\bibnamefont {Wulff}}, \bibinfo {author}
  {\bibfnamefont {L.}~\bibnamefont {Wernli}}, \bibinfo {author} {\bibfnamefont
  {A.}~\bibnamefont {Copetudo}}, \bibinfo {author} {\bibfnamefont
  {D.}~\bibnamefont {Malz}}, \bibinfo {author} {\bibfnamefont {P.}~\bibnamefont
  {Magnard}}, \bibinfo {author} {\bibfnamefont {A.}~\bibnamefont {Akin}},
  \bibinfo {author} {\bibfnamefont {M.}~\bibnamefont {Gabureac}}, \bibinfo
  {author} {\bibfnamefont {G.~J.}\ \bibnamefont {Norris}}, \bibinfo {author}
  {\bibfnamefont {J.~I.}\ \bibnamefont {Cirac}}, \bibinfo {author}
  {\bibfnamefont {A.}~\bibnamefont {Wallraff}},\ and\ \bibinfo {author}
  {\bibfnamefont {C.}~\bibnamefont {Eichler}},\ }\bibfield  {title} {\bibinfo
  {title} {Realizing a deterministic source of multipartite-entangled photonic
  qubits},\ }\href {https://doi.org/10.1038/s41467-020-18635-x} {\bibfield
  {journal} {\bibinfo  {journal} {Nature Communications}\ }\textbf {\bibinfo
  {volume} {11}},\ \bibinfo {pages} {4877} (\bibinfo {year}
  {2020})}\BibitemShut {NoStop}%
\bibitem [{\citenamefont {Wan}\ \emph {et~al.}(2021)\citenamefont {Wan},
  \citenamefont {Choi}, \citenamefont {Kim}, \citenamefont {Shutty},\ and\
  \citenamefont {Hayden}}]{Wan2021PRX}%
  \BibitemOpen
  \bibfield  {author} {\bibinfo {author} {\bibfnamefont {K.}~\bibnamefont
  {Wan}}, \bibinfo {author} {\bibfnamefont {S.}~\bibnamefont {Choi}}, \bibinfo
  {author} {\bibfnamefont {I.~H.}\ \bibnamefont {Kim}}, \bibinfo {author}
  {\bibfnamefont {N.}~\bibnamefont {Shutty}},\ and\ \bibinfo {author}
  {\bibfnamefont {P.}~\bibnamefont {Hayden}},\ }\bibfield  {title} {\bibinfo
  {title} {Fault-tolerant qubit from a constant number of components},\ }\href
  {https://doi.org/10.1103/PRXQuantum.2.040345} {\bibfield  {journal} {\bibinfo
   {journal} {PRX Quantum}\ }\textbf {\bibinfo {volume} {2}},\ \bibinfo {pages}
  {040345} (\bibinfo {year} {2021})}\BibitemShut {NoStop}%
\bibitem [{\citenamefont {Thomas}\ \emph {et~al.}(2024)\citenamefont {Thomas},
  \citenamefont {Ruscio}, \citenamefont {Morin},\ and\ \citenamefont
  {Rempe}}]{Thomas2024}%
  \BibitemOpen
  \bibfield  {author} {\bibinfo {author} {\bibfnamefont {P.}~\bibnamefont
  {Thomas}}, \bibinfo {author} {\bibfnamefont {L.}~\bibnamefont {Ruscio}},
  \bibinfo {author} {\bibfnamefont {O.}~\bibnamefont {Morin}},\ and\ \bibinfo
  {author} {\bibfnamefont {G.}~\bibnamefont {Rempe}},\ }\bibfield  {title}
  {\bibinfo {title} {Fusion of deterministically generated photonic graph
  states},\ }\href {https://doi.org/10.1038/s41586-024-07357-5} {\bibfield
  {journal} {\bibinfo  {journal} {Nature}\ }\textbf {\bibinfo {volume} {629}},\
  \bibinfo {pages} {567} (\bibinfo {year} {2024})}\BibitemShut {NoStop}%
\bibitem [{\citenamefont {Li}\ \emph {et~al.}(2022)\citenamefont {Li},
  \citenamefont {Economou},\ and\ \citenamefont {Barnes}}]{Li2022}%
  \BibitemOpen
  \bibfield  {author} {\bibinfo {author} {\bibfnamefont {B.}~\bibnamefont
  {Li}}, \bibinfo {author} {\bibfnamefont {S.~E.}\ \bibnamefont {Economou}},\
  and\ \bibinfo {author} {\bibfnamefont {E.}~\bibnamefont {Barnes}},\
  }\bibfield  {title} {\bibinfo {title} {Photonic resource state generation
  from a minimal number of quantum emitters},\ }\href
  {https://doi.org/10.1038/s41534-022-00522-6} {\bibfield  {journal} {\bibinfo
  {journal} {npj Quantum Information}\ }\textbf {\bibinfo {volume} {8}},\
  \bibinfo {pages} {11} (\bibinfo {year} {2022})}\BibitemShut {NoStop}%
\bibitem [{\citenamefont {Aaronson}\ and\ \citenamefont
  {Gottesman}(2004)}]{Aaronson2004PRA}%
  \BibitemOpen
  \bibfield  {author} {\bibinfo {author} {\bibfnamefont {S.}~\bibnamefont
  {Aaronson}}\ and\ \bibinfo {author} {\bibfnamefont {D.}~\bibnamefont
  {Gottesman}},\ }\bibfield  {title} {\bibinfo {title} {Improved simulation of
  stabilizer circuits},\ }\href {https://doi.org/10.1103/PhysRevA.70.052328}
  {\bibfield  {journal} {\bibinfo  {journal} {Phys. Rev. A}\ }\textbf {\bibinfo
  {volume} {70}},\ \bibinfo {pages} {052328} (\bibinfo {year}
  {2004})}\BibitemShut {NoStop}%
\end{thebibliography}
\end{document}